\documentclass[manuscript,screen]{acmart}
\AtBeginDocument{%
  }

\usepackage{comment}
\usepackage{csquotes}
\usepackage[nohyperlinks]{acronym}
\usepackage{cleveref}
\usepackage{graphicx}
\usepackage{placeins}
\usepackage[countmax]{subfloat}
\usepackage{subcaption}
\usepackage[inline]{enumitem}
\usepackage{tcolorbox}
\usepackage{microtype}
\usepackage{wrapfig}
\usepackage{makecell}
\usepackage{listings}
\usepackage{tikz}
\usetikzlibrary{positioning}
\acrodef{ai}[AI]{Artificial Intelligence}
\acrodef{cot}[CoT]{Chain-of-Thought}
\acrodef{dnn}[DNN]{Deep Neural Network}
\acrodefplural{dnn}[DNNs]{Deep Neural Networks}
\acrodef{dsl}[DSL]{Domain Specific Language}
\acrodef{genai}[GenAI]{Generative Artificial Intelligence}
\acrodef{hci}[HCI]{Human-Computer Interaction}
\acrodef{llm}[LLM]{Large Language Model}
\acrodefplural{llm}[LLMs]{Large Language Models}
\acrodef{lsl}[LSL]{LLM Scripting Language}
\acrodef{rag}[RAG]{Retrieval Augmented Generation}
\acrodef{repl}[REPL]{Read–eval–print loop}
\acrodef{se}[SE]{Software Engineering}
\acrodef{xai}[XAI]{Explainable AI}
\acrodef{nlo}[NLO]{Natural Language Outlines}

\setcopyright{acmlicensed}
\copyrightyear{2025}
\acmYear{2025}
\acmDOI{XXXXXXX.XXXXXXX}
\acmISBN{978-1-4503-XXXX-X/2026/01}

\begin{document}

\title{A Roadmap for Tamed Interactions with Large Language Models }

\author{Vincenzo Scotti}
\email{vincenzo.scotti@kit.edu}
\orcid{0000-0002-8765-604X}
\affiliation{%
  \institution{Karlsruhe Institute of Technology (KIT)}
  \city{Karlsruhe}
  \country{Germany}
}

\author{Jan Keim}
\email{jan.keim@kit.edu}
\orcid{0000-0002-8899-7081}
\affiliation{%
  \institution{Karlsruhe Institute of Technology (KIT)}
  \city{Karlsruhe}
  \country{Germany}
}

\author{Tobias Hey}
\email{hey@kit.edu}
\orcid{0000-0003-0381-1020}
\affiliation{%
  \institution{Karlsruhe Institute of Technology (KIT)}
  \city{Karlsruhe}
  \country{Germany}
}

\author{Andreas Metzger}
\email{andreas.metzger@paluno.uni-due.de}
\orcid{0000-0002-4808-8297}
\affiliation{%
  \institution{Ruhr Institute for Software Technology (paluno), University of Duisburg-Essen}
  \city{Essen}
  \country{Germany}
}

\author{Anne Koziolek}
\email{koziolek@kit.edu}
\orcid{0000-0002-1593-3394}
\affiliation{%
  \institution{Karlsruhe Institute of Technology (KIT)}
  \city{Karlsruhe}
  \country{Germany}
}

\author{Raffaela Mirandola}
\email{raffaela.mirandola@kit.edu}
\orcid{0000-0003-3154-2438}
\affiliation{%
  \institution{Karlsruhe Institute of Technology (KIT)}
  \city{Karlsruhe}
  \country{Germany}
}

\renewcommand{\shortauthors}{Scotti et al.}

\begin{abstract}

\acp{llm} are increasingly embedded in software systems (\acs{genai}ware), enabling new forms of automation and interaction. 
However, their probabilistic nature and reliance on prompt programming challenge reliability, robustness, and maintainability. 
In current practice, prompt-related concerns (e.g., context management, interaction logic, output validation) are embedded in general-purpose code, leading to implicit, hard-to-analyze systems.
We argue that prompt programming should be treated as a first-class \ac{se} concern and propose \ac{lsl}, a \ac{dsl} for structuring \ac{llm} interactions as analyzable programs. 
\ac{lsl} introduces abstractions for interaction blocks, context scopes, output constraints, and control flow, separating deterministic logic from probabilistic model behavior while ensuring syntactic compliance.
From an \ac{se} perspective, \ac{lsl} supports disciplined development by making interaction logic explicit, analyzable, and amenable to verification and validation. 
It also acts as cognitive scaffolding, externalizing prompt design into programmable artifacts that reduce implicit reasoning and support systematic debugging, evolution, and reuse.
We illustrate these properties in a structured generation scenario, showing improved failure localization and interaction transparency. 
While \ac{lsl} does not guarantee semantic correctness or factual accuracy, it provides a principled foundation for more analyzable and maintainable prompt-based systems.

\end{abstract}

\acresetall{}




\maketitle

\section{Introduction}

The advancements in \ac{genai}~\cite{llm_se_roadmap}, and in particular \acp{llm}~\cite{llm_survey}, are rapidly reshaping how \ac{se} addresses software systems design and development.
Beyond their role as development-time assistants or copilots~\cite{github2023copilot}, \acp{llm} are increasingly embedded directly into software products to realize functionality that would otherwise be infeasible or too expensive to implement~\cite{llm_se_roadmap}.
In such \emph{\ac{genai}ware} systems~\cite{llm_se_roadmap}, foundation models are invoked at runtime as integral software components, powering capabilities such as conversational interfaces, information retrieval, content generation, and adaptive decision support~\cite{DBLP:journals/corr/abs-2406-10300}.
As a result, software systems are no longer defined solely by code and data, but also by \emph{prompts} (i.e., natural language instructions) that govern the behavior of these models.
This shift introduces new \ac{se} challenges, as developers must now reason about artifacts that are inherently non-deterministic, difficult to validate, and hard to maintain.
In practice, this leads to issues such as inconsistencies in generated outputs, propagation of invalid results, and increased difficulty in debugging and evolving \ac{llm}-based systems~\cite{10.1145/3806396,DBLP:journals/tosem/ChenGCZL25,DBLP:conf/icsm/AbbassiSNK25}.

Recent research has observed that prompts in \ac{genai}ware are no longer simple textual inputs, but are becoming meticulously engineered parts of the codebase, effectively functioning as programs themselves~\cite{DBLP:journals/pacmse/LiangLRM25,llm_se_roadmap}.
These prompts encode \emph{task definitions}, \emph{interaction protocols}, \emph{constraints}, and \emph{expectations about model outputs}, and they critically shape the behavior of the overall system.
More broadly, \ac{genai}ware development increasingly involves context engineering~\cite{DBLP:journals/corr/abs-2507-13334}, which consists in the construction, scoping, and evolution of the information provided to an \ac{llm}, and output constraining~\cite{DBLP:conf/emnlp/GengJP023,DBLP:conf/nips/ParkWBPD24}, which consists of specifying structural and contractual expectations on generated results (either to validate the output or to enforce specific behaviour).
For simplicity, we refer to this combination of prompt design, context management, and output constraints as \emph{prompt programming}.

Together, the prompt programming components determine how foundation models behave within software workflows.
Recent work in \ac{hci} further reinforces this view by showing that prompt programming exhibits properties analogous to software development, despite lacking a formal grammar or programming language structure~\cite{10.1145/3544549.3585737}.
However, despite their importance, these elements are still largely treated as ad-hoc strings or implicit conventions, embedded in general-purpose code and refined through trial and error.
This practice makes these prompt-based systems fragile, hard to debug, difficult to evolve, and challenging to reason about from a \ac{se} perspective~\cite{llm_se_roadmap}.

This situation exposes a fundamental \ac{se} gap, and the root of the problem is not expressiveness but \emph{analyzability}.
Existing languages and frameworks are expressive enough: prompt templates can be written in Python, output validators can be defined with libraries such as Pydantic, and orchestration pipelines can be assembled with frameworks such as LangChain.
What these approaches cannot provide is a clear separation between deterministic software logic and probabilistic model behavior.
When prompt programming is embedded in general-purpose code, interaction logic becomes implicit, verification targets become unclear, and static analysis is practically infeasible, because conventional control flow is interleaved with probabilistic \ac{llm} invocations in ways the host language cannot reason about.

In \ac{genai}ware systems, developers therefore face increased difficulty in inspecting the structure of interactions with \acp{llm}, enforcing invariants on generated outputs, and reasoning about the behavior of prompt-driven workflows over time.
As systems grow in complexity, this lack of analyzable structure undermines reliability, maintainability, and trustworthiness, and prevents the systematic application of good \ac{se} practices such as verification, validation, and lifecycle management.

In this paper, we argue that addressing these challenges requires treating prompt programming as a \emph{first-class} \ac{se} concern, and that doing so demands dedicated programming abstractions, not merely libraries, that make prompt behavior
\begin{enumerate*}[label=(\roman*)]
    \item \emph{separable}: the LLM interactions (prompts, scaffolding code etc.) should be clearly separated from code that is not related to prompt programming;  
  \item \emph{inspectable}: the structure and intent of each interaction is explicit and human-readable;
  \item \emph{analyzable}: static and runtime analysis can reason about control flow, reachability, and contracts; and
  \item \emph{governable}: output constraints and execution invariants are part of the language semantics, not bolted on as external validators.
\end{enumerate*}
As such, we propose the concept of \ac{lsl}: a \ac{dsl} designed to express prompt-based interactions as structured programs.
Rather than replacing general-purpose programming languages or existing \ac{genai} frameworks, \ac{lsl} defines a \emph{restricted interaction layer} that sits between deterministic software and probabilistic foundation models.
In this sense, \ac{lsl} is intended to play a role analogous to SQL for databases or regular expressions for pattern matching: it does not supersede host languages, but provides a specialized abstraction that makes reasoning, optimization, and verification amenable in ways that would be impractical in unrestricted general-purpose code.

\ac{lsl} is intended to define prompts as executable specifications.
It provides constructs to structure interactions with \acp{llm}, manage and scope context, constrain outputs, and orchestrate multi-step workflows involving both model invocations and conventional software components.
By making prompt logic explicit and declarative, \ac{lsl} enables reasoning about the structure and control flow of \ac{genai}ware interactions, and opens the door to applying established \ac{se} techniques (such as static analysis, runtime validation, testing, and monitoring) at the level of prompt programs.
While, \ac{lsl} does not aim to guarantee semantic correctness or factual truthfulness of model outputs, it offers structural guarantees about interactions, contracts, and execution paths, thereby improving engineering control over prompt-driven systems.

The novelty of this work lies exactly in reframing prompt programming into a first-class programming model. 
While existing frameworks provide mechanisms for prompt templating, structured outputs, and workflow orchestration, they treat these concerns as library-level constructs embedded in general-purpose code. 
In contrast, \ac{lsl} elevates interactions with the \ac{llm} to explicit, analyzable program artifacts, where context management, output constraints, and control flow are integrated into the language semantics. 
This enables forms of reasoning—such as static analysis of interaction structure, verification of workflow properties, and systematic enforcement of interaction contracts—that are not achievable when prompt logic remains implicit within host languages.
In this sense, \ac{lsl} is not a convenience layer over existing tooling, but a shift in abstraction boundary from ``using \acp{llm} inside programs'' to ``programming interactions with \acp{llm} as programs''.

From a human–\ac{ai} interaction perspective, \ac{lsl} follows a mixed-initiative approach.
Pure natural language is flexible but semantically unstable across executions; pure code is rigorous but too rigid to concisely express domain intent directed at a probabilistic model.
\ac{lsl} occupies the productive middle ground: natural language remains central for expressing task intent and domain knowledge, while lightweight formal structure stabilizes and constrains that intent, making it amenable to analysis and enforcement.

This paper presents a vision and research roadmap for \ac{lsl} as a prompt-programming language for \ac{genai}ware.
We position \ac{lsl} within the broader landscape of \ac{genai}ware development, clarify its scope and intended guarantees, and outline how it can support reliability, robustness, and trustworthiness in prompt-based systems.
By reframing prompts as programs and treating prompt programming as a core \ac{se} concern, we aim to contribute to the foundations needed to engineer \ac{genai}ware systems with the same discipline expected of traditional software.

We divide the rest of this vision paper into the following sections.
In \Cref{sec:background}, we recap the essentials about \acp{llm}.
In \Cref{sec:sota}, we summarise the main work related to ours and position our contribution.
In \Cref{sec:llmscriptinglanguage}, we describe \ac{lsl} in detail.
In \Cref{sec:illustrative-example}, we present an application example to illustrate the use of \ac{lsl}.
In \Cref{sec:opportunities} and \Cref{sec:challenges}, we comment, respectively, on the future opportunities and challenges we foresee.
In \Cref{sec:evaluation} we report the design and results of our evaluation investigating the need of abstractions like \ac{lsl}.
In \Cref{sec:roadmap}, we present a research roadmap towards realizing the vision of \ac{lsl}.
Finally, in \Cref{sec:conclusion}, we summarize our considerations.

\section{Background}
\label{sec:background}

In this section, we provide an overview of \acp{llm} (\Cref{sec:overview}), highlighting capabilities (\Cref{sec:capabilities}) and limitations (\Cref{sec:limitations}).

\subsection{Overview}
\label{sec:overview}

\acp{llm} are \acp{dnn} trained on massive amounts of data to learn a probabilistic generative model of text.
The underlying \ac{dnn} of a \ac{llm} is a Transformer~\cite{DBLP:conf/nips/VaswaniSPUJGKP17}, a \acp{dnn} designed to process sequential data, like sequences of text tokens\footnote{Tokens are the basic units processed by a \ac{llm} (e.g., words, sub-words, or characters).}, through the \emph{self-attention mechanisms}~\cite{DBLP:conf/nips/VaswaniSPUJGKP17}.
This Transformer architecture enables \acp{llm} to capture long and complex dependencies throughout the input text as well as memorise and recall a large amount of information, allowing the generation of highly coherent and contextually relevant text.

We mainly distinguish these \acp{llm} based on their size, computed in the number of parameters of the underlying \ac{dnn}, since, usually, their capabilities depend on the size of the model and the amount of training data.
Bigger and more capable models, including closed-access models such as \emph{GPT} or \emph{ChatGPT}~\cite{DBLP:journals/corr/abs-2303-08774,DBLP:journals/corr/abs-2410-21276} and \emph{Gemini}~\cite{DBLP:journals/corr/abs-2312-11805,DBLP:journals/corr/abs-2403-05530}, are hundreds or thousands of billions of parameters large.
Smaller models, on the other hand, have only up to tens of billions of parameters.
These include open-access models such as \emph{DeepSeek}~\cite{DBLP:journals/corr/abs-2401-02954,DBLP:journals/corr/abs-2405-04434,DBLP:journals/corr/abs-2412-19437}, \emph{Llama}~\cite{DBLP:journals/corr/abs-2307-09288,DBLP:journals/corr/abs-2407-21783}, \emph{Vicuna}~\cite{lmsys2023vicuna}, \emph{Mistral}~\cite{DBLP:journals/corr/abs-2310-06825,DBLP:journals/corr/abs-2401-04088}, and \emph{Gemma}~\cite{DBLP:journals/corr/abs-2403-08295,DBLP:journals/corr/abs-2408-00118}.
The latter group represents a set of highly valuable alternatives for research and application development that, under proper tweaking, can rival the bigger ones.
Moreover, these open models are often also released in larger versions~\cite{DBLP:journals/corr/abs-2412-19437,DBLP:journals/corr/abs-2407-21783,DBLP:journals/corr/abs-2401-04088}.

Open-access \acp{llm} are mainly distributed, sharing the neural network weights learned during the pre-training.
Often, to make these pre-trained models usable as assistants (like ChatGPT), they are \emph{fine-tuned} (i.e., refined via further training) to behave as \emph{instruction-following} agents or \emph{chatbot-assistants}~\cite{DBLP:conf/iclr/SanhWRBSACSRDBX22,DBLP:journals/corr/abs-2210-11416,DBLP:journals/csur/ScottiST24} by training on conversations (or other interactions) where a user and an agent communicate via natural language to solve some tasks.
Typically, the fine-tuning data is formatted using a specific template, where the input is a sequence composed of:
\emph{system message} (main task instructions, which can also be chatbot directives);
\emph{user message} (user's input to be processed, question, or request);
\emph{agent response} (agent's output or response to user's question or request).
Each sample in the data set contains one or more message-response exchanges between the user and the agent (\ac{llm}) covering multiple tasks, multiple steps within a task, or to handle user corrections.
Alternatively, other models are distributed as coding assistants, like in the case of \emph{Copilot}~\cite{github2023copilot} or \emph{Code Llama}~\cite{DBLP:journals/corr/abs-2308-12950}, where fine-tuning is used to refine the model on source code generation, which is, however, a capability most pre-trained or assistant \acp{llm} already have.
Differently from the past years, where fine-tuning was used to adapt a pre-trained model to a very specific task like classification (often modifying the architecture)~\cite{radford2018improving,DBLP:conf/naacl/DevlinCLT19}, nowadays prompting (i.e., instructing the model via natural language) is preferred to solve most problems with \acp{llm} and fine-tuning is applied very rarely. %

\subsection{Capabilities}
\label{sec:capabilities}

\begin{figure}[!ht]
\begin{center}
    \subfloat[Legend \label{fig:llm_legend}]{\includegraphics[width=.729\columnwidth]{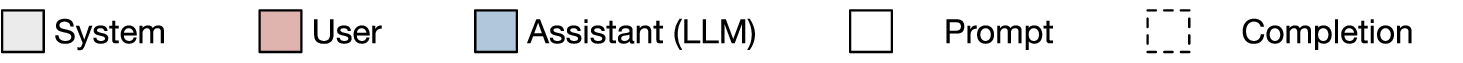}} \\
    \subfloat[Few-shot learning \label{fig:llm_fsl}]{\includegraphics[width=.24075\columnwidth]{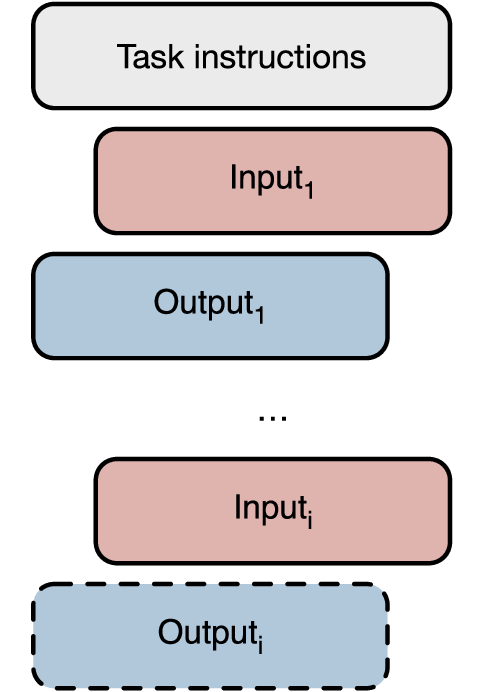}} \hfill
    \subfloat[RAG\label{fig:llm_rag}]{\includegraphics[width=.24075\columnwidth]{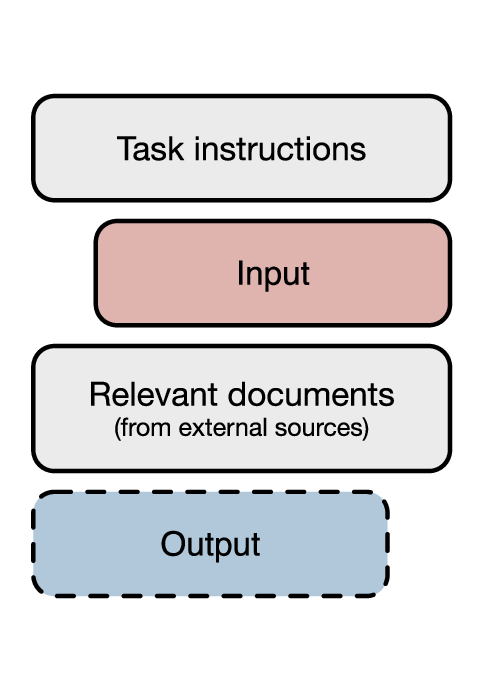}} \hfill
    \subfloat[CoT reasoning \label{fig:llm_cot}]{\includegraphics[width=.24075\columnwidth]{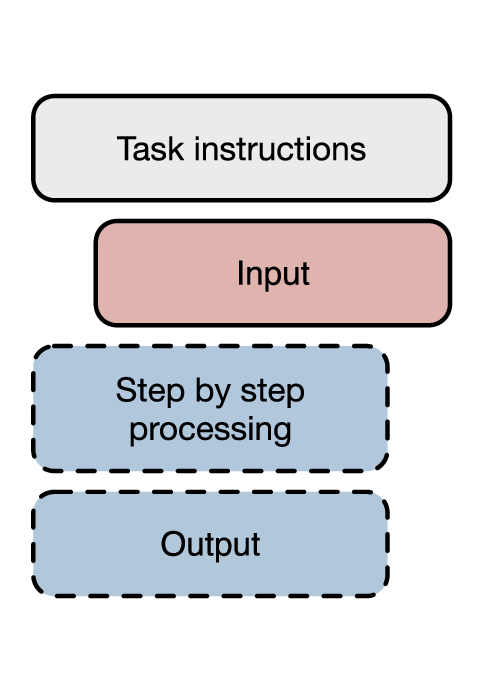}} %
\caption{\ac{llm} prompting approaches.}
\label{fig:llm}
\end{center}
\end{figure}

Auto-regressive (or causal) \acp{llm}, which represent the vast majority of current models, are designed to generate text autoregressively.
They yield the probability distribution from which to sample the next token one at a time, conditioning the distribution of this next token on all the preceding ones (the context).
From a practical perspective, \acp{llm} are used by prompting with some prefix text, which may include some instructions for a task or some user input, and generating an output completion by sampling iteratively from the predicted probability distribution.

An \ac{llm} can be prompted to tackle problems in different ways.
The main prompting approaches, which we report in \Cref{fig:llm}, are:
\emph{few-shot learning} (sometimes called in-context learning)~\cite{DBLP:conf/nips/BrownMRSKDNSSAA20},
\emph{Retrieval Augmented Generation} (RAG)~\cite{DBLP:journals/corr/abs-2312-10997}, and
\emph{Chain of Thought} (CoT) \emph{reasoning}~\cite{DBLP:conf/nips/KojimaGRMI22}.
Few-shot learning, depicted in \Cref{fig:llm_fsl}, consists of providing additional examples of input-output pairs (the shots) together with the task instructions as part of the prompt, helping the \ac{llm} better \enquote{understand} the task and the expected output format.
This is an example of the ability of \acp{llm} to generalize from limited data.
RAG, depicted in \Cref{fig:llm_rag}, consists of automatically extending the input context by integrating information coming from external knowledge sources and, thus, allowing the model to ground its output responses in this relevant information.
CoT reasoning, depicted in \Cref{fig:llm_cot}, consists of forcing the model to generate a step-by-step reasoning process before the actual output, giving the model more space for its internal \enquote{reasoning} processes and, thus, improving the performance on complex tasks.

All these typical approaches are orthogonal to each other, meaning that they can be combined for better performance.
Moreover, none of them is specific to the instruction-following or chatbot assistant fine-tuning and can be used with pre-trained models~\cite{DBLP:conf/nips/BrownMRSKDNSSAA20}.

Beyond their role as passive text generators, \acp{llm} increasingly support deployment in \emph{agentic} configurations, where they operate as (semi-)autonomous decision‑makers~\cite{DBLP:journals/access/AcharyaKB25}. 
In such settings, a \ac{llm} is prompted with explicit goals, contextual system information, and access to external tools such as search, simulation, or code execution. 

\subsection{Limitations}
\label{sec:limitations}

\acp{llm} come with very general capabilities and high flexibility, but they are prone to many known issues.
These issues range from architectural limitations to problems with the generated content.

From the \ac{dnn} architecture side, we identify two main issues:
context size (i.e., the maximum number of consecutive tokens a model can manage) and
computational complexity.
Even though in the last few years we have seen an increase in techniques designed to extend physically or virtually the maximum context, the problem of having a finite context still remains, and, especially in smaller models, the attention doesn't keep up well when the context grows excessively.
Moreover, the computation of the \emph{self-attention mechanism} (core of the Transformer) is quadratic in time to the length of the input.
There are caching mechanisms and alternative implementations~\cite{DBLP:journals/corr/abs-2112-05682,DBLP:conf/nips/DaoFERR22,DBLP:conf/iclr/Dao24} to cope with these issues, but they aren't always usable.
Besides their architectural limitations, \acp{llm} stay probabilistic models.
This nature makes them robust to noise and uncertainty, but also prone to errors, as they rely on probability rather than sound values.
These errors can be caused by different factors, including sensitivity to input variations or ambiguities, and overfitting to training knowledge and patterns.

As a result of the technical issues, the models end up relying on more recent context and may end up generating contradictions, struggling with multi-step reasoning, or simply deviating from the original context and instructions.
Moreover, due to the probabilistic nature, it may end up generating biased content based on unreliable, outdated, or even unethical information.
At the same time, this probabilistic nature may lead to similar phenomena called hallucinations.
An \ac{llm} creates hallucinations to simply generate a completion that is more statistically likely, but the output consists of made-up facts and details that are neither grounded in world knowledge, causing problems with factuality~\cite{DBLP:journals/csur/JiLFYSXIBMF23}, nor in the current context, causing problems with faithfulness~\cite{DBLP:journals/csur/JiLFYSXIBMF23}.

The combination of these issues makes these models unsuitable for tasks that require high reliability, robustness, and trustworthiness, like automating a pipeline using the \ac{llm}.
Consider the case where we are interested in interaction with a user and generate a call to some external API, and the model fails to close a parenthesis.
The use of techniques like few-shot learning may help the model to stick to a given pattern (e.g., generate well-formed JSON).
Similarly, \ac{rag} helps to integrate more updated and reliable knowledge (e.g., refer to the latest news), and \ac{cot} reasoning may help improve accuracy in multi-step reasoning tasks (e.g., solving an equation).
Yet, all these techniques are nothing but palliatives as they only help to reduce the probability that an error occurs, but without any guarantee that the problem won't show up.

\section{Related Work}
\label{sec:sota}

In this section, we present the state of the art of \ac{se} support for \ac{genai}ware (\Cref{sec:sephorai}), then, we focus on the existing frameworks designed to manage the interaction with \acp{llm} (\Cref{sec:existingframeworks}), and on the known challenges arising from \ac{genai}ware development, and, finally, we position the proposed \ac{lsl} framework with respect to the state-of-the-art (\Cref{sec:positioningofthiswork}).

\subsection{SE for AI}
\label{sec:sephorai}

This paper elaborates on how the introduction of \ac{ai}-powered components within software systems (\ac{genai}ware), particularly those relying on \acp{llm}, should be paired with appropriate tools for the development of such software systems that allow for the correct use of such \ac{ai} components.
As mentioned above, realising this potential at scale requires a paradigm shift in how these applications are developed, deployed, and maintained.
Yet, current practices, often characterised by ad-hoc prompt engineering and fragmented toolchains, hinder productivity, reproducibility, and the development of robust, enterprise-grade \ac{genai}ware~\cite{DBLP:conf/cain/BarnettKTB024,DBLP:conf/icse-seip/NaharKBPZB25,DBLP:conf/icse/Shao0S00025}.
Empirical studies show that developers struggle to engineer \ac{genai}ware in practice, with low resolution rates for prompt-related issues and long debugging cycles, highlighting the lack of appropriate abstractions for structuring interactions~\cite{DBLP:journals/tosem/ChenGCZL25,DBLP:journals/corr/abs-2503-02400,DBLP:journals/pacmse/LiangLRM25}.

A range of tools and approaches has been proposed to facilitate the development of AI-based software systems, including libraries for prompt templating, retrieval tools that integrate semantic search, and plug-ins for AI-assisted coding. These tools address specific aspects of AI integration, such as handling prompts~\cite{DBLP:journals/corr/abs-2503-02400,DBLP:journals/pacmse/LiangLRM25}, leveraging curated data through techniques like RAG~\cite{DBLP:conf/cain/BarnettKTB024,DBLP:conf/icse/Shao0S00025}, and embedding AI capabilities within traditional software workflows. However, existing solutions often treat these components in isolation. In contrast, our approach seeks to provide a more unified framework that considers \emph{prompts} and \emph{data}, alongside \emph{code}, as first-class citizens in the software development process.

In the 80s, the concept of \emph{literate programming}, where computers could be instructed purely through natural language, seemed science fiction~\cite{DBLP:journals/cj/Knuth84}.
Yet, we are now moving in that direction: tools like \ac{nlo} offer a novel modality and interaction surface for providing \ac{ai} assistance to developers throughout the software development process~\cite{DBLP:conf/sigsoft/ShiAACGKNPRRRST25}.
Such \ac{nlo} enable a bidirectional sync between code and natural language, where a developer can change either code or NL and have the LLM automatically update the other.
In contrast, \ac{lsl} tries to go beyond this iterative approach managed by an assistant and tries to mix together structured code with natural language, so that where necessary the \ac{ai} component can be easily ``programmed'' using a prompt.
Moreover, the suite coming with \ac{lsl} is thought to integrate also tools for verification and validation, rather than the ``bare'' programming components.

While \acp{llm} are increasingly used in \ac{se}, there is still no fully integrated approach that embeds them across the DevOps lifecycle.
Software engineers are dealing with the following problems:
\begin{enumerate*}
    \item \emph{Fragmented tooling}: current tools are often disparate, requiring developers to stitch together multiple components, leading to inefficiencies and compatibility issues~\cite{DBLP:conf/icse-seip/NaharKBPZB25}.
    \item \emph{Lack of standardization}: there's a dearth of industry-wide standards for prompt engineering, promptware architecture, interfaces, and lifecycle management, leading to inconsistent development practices~\cite{DBLP:journals/pacmse/LiangLRM25,lintang_sutawika_2024_14506035,DBLP:conf/icse-seip/NaharKBPZB25}.
    \item \emph{Limited verification and debugging capabilities}: debugging \ac{llm}-based applications, especially hallucination-related issues, remains challenging due to their black-box nature and probabilistic outputs~\cite{DBLP:conf/icse-seip/NaharKBPZB25,DBLP:journals/corr/abs-2503-00481}.
    \item \emph{Immature lifecycle management}: While \emph{MLOps} addresses model lifecycle~\cite{10.1145/3747346}, the specific nuances of prompt and data versioning, testing, and deployment within a combined software-prompt-data paradigm are not fully mature~\cite{DBLP:conf/icse-seip/NaharKBPZB25}.
\end{enumerate*}

Problems like the fragmented tooling (1) and the lack of standardisation (2) are not new for \ac{se} research.
In fact, there were similar issues when several cloud providers started appearing, which led to the development of multi-cloud research to help deal with the different interfaces exposed by the vendors~\cite{DBLP:conf/ecsa/BrogiCCDNGPPS16}.

Nevertheless, the \ac{se} community is moving to improve several aspects related to the development of \ac{genai}ware. %
Practice to take \ac{ai} components into account now starts from the early requirement elicitation stage~\cite{10771003} and spans throughout the entire life cycle processes~\cite{ISOIEC5338}.
Prompt engineering and prompt programming are currently evolving to produce robust solutions that can keep up with the ever-changing nature of \acp{llm}, also becoming integral components of software applications~\cite{DBLP:journals/corr/abs-2503-02400,DBLP:journals/pacmse/LiangLRM25}, also known as \emph{promptware}.
Data management is strongly influencing the development as well, leading to architectures specifically designed for properly integrating \acp{llm} and data within the software system~\cite{DBLP:conf/cain/BarnettKTB024,DBLP:conf/icse/Shao0S00025}.
Recent work has explored IDE-inspired tooling for prompt programming, including semantic highlighting, structured editing, and automated detection of prompt structure~\cite{10.1145/3544549.3585737}.
In contrast to such tool-centric approaches, our work focuses on providing language-level abstractions that make interaction structure explicit and analyzable.

Beyond engineering practices, there are also emerging efforts to rethink programming models for \ac{genai}ware at a more fundamental level. One example is Oracular Programming~\cite{oracular-programming-2025}, which proposes a paradigm where \acp{llm} act as runtime oracles that resolve high-level nondeterministic strategies. This approach emphasizes modularity, evolvability, and extensibility, treating calls to \acp{llm} as structured programming constructs rather than ad-hoc prompts. While our proposal of \ac{lsl} focuses on scripting, constraining, and verifying interactions to provide stronger guarantees, Oracular Programming highlights a complementary research direction in which language abstractions are designed to embrace nondeterminism while still supporting systematic reasoning about \ac{genai}ware systems.

\subsection{Existing Frameworks}
\label{sec:existingframeworks}

With our work, we are interested in covering and unifying several aspects related to the adoption of \acp{llm} and \ac{llm}-powered components within software systems.
There are several key aspects we are interested in.
First, we want to look into application development, looking into prompt management, function calling, agentic AI, and data management.
Second, we want to improve reliability and robustness through standard interfaces, automatic input templating and formatting, and output constraints specification. Third, we want to create trustworthiness through verification and validation, and through explainability.

Looking at the state of the art in terms of tools and frameworks addressing these issues, we could not help but notice a strong fragmentation and a lack of standardisation that are preventing the automation of many aspects of \ac{ai}-powered software life cycle.

About application development frameworks and tools to offer the different functionalities, \emph{LangChain} and \emph{LangGraph}~\cite{docs/introduction/} are the earliest and most famous examples.
Yet, through the last years alternatives like \emph{Semantic Kernel}~\cite{https://github.com/microsoft/semantic-kernel}, \emph{AutoGen} ~\cite{DBLP:journals/corr/abs-2308-08155}, \emph{Flowise} ~\cite{https://github.com/FlowiseAI/Flowise}.
Some of these frameworks, like \emph{CrewAI}~\cite{https://github.com/crewAIInc/crewA}, are specifically designed for agentic or multi-agent application development.
The most recent contribution in this sense is the \ac{nlo} tool~\cite{DBLP:conf/sigsoft/ShiAACGKNPRRRST25}, which brings the development closer to literate programming.
Unlike inference engines, many solutions for context management attempt to provide a unified interface to all models; however, they often lack the ability to constrain the generated output.

On the data side of context management, in recent years, frameworks like \emph{Llama Index}~\cite{https://github.com/run-llama/llama_index} have helped automate most of the work involved in building a vector document store.
These frameworks allow to enhance data-intensive applications requiring \ac{rag} to offer their functionalities.
In this sense, \emph{Haystack}~\cite{https://github.com/deepset-ai/haystack} represents a valuable tool that provides resources for both data management and models orchestration.

With a specific focus on reliability and robustness, we mostly depend on inference engines or frameworks.
Solutions like \emph{Llama CPP} and \emph{Llama CPP Python}~\cite{llama.cpp,llama.cpp.python} offers many low-level functionalities for templating the input and capturing elements of the output.
Unlike other existing inference engines and tools, Llama CPP allows specifying a \emph{generative grammar} to constrain the generated completions.
We talk in these cases of \emph{grammar-constrained decoding}~\cite{DBLP:conf/emnlp/GengJP023}.
The closest tool in this sense is \emph{Guidance}~\cite{guidance-ai/guidance} that offers a templating system well blended into Python syntax and the possibility to specify capture groups to be automatically applied to the output and to specify a finite list of alternative completions to select from.
However, most of these tools rely only on HuggingFace's Transformers API or use their own custom interfaces.
Prior work shows that \ac{llm}-generated artifacts frequently contain structural defects, including malformed code, incomplete outputs, and inconsistencies across steps, even when they are functionally plausible~\cite{DBLP:conf/icsm/AbbassiSNK25}.

There are some tools specialised in output constraints and validation.
\emph{NeMo Guardrails}~\cite{DBLP:conf/emnlp/RebedeaDSPC23} was designed to impose programmable guardrails to make sure a dialogue with an assistant follows a specific template or that certain harmful topics are avoided.
Many of these tools are thought to impose safety guardrails~\cite{DBLP:journals/corr/abs-2312-06674}, and in fact, there are specialised solutions to capture or prevent toxic and harmful content generation, like \emph{Guardrails AI} ~\cite{https://www.guardrailsai.com/}.

Concerning trustworthiness, we have two main aspects being touched: verification and validation on one side, and explainability on the other.
About verification and validation, we have that now testing tools and frameworks like \emph{LangTest}~\cite{DBLP:journals/simpa/NazirCCKSMKT24} or \emph{EvoTox}~\cite{DBLP:journals/tse/CorboBDLSC25} which are designed to take into account the black-box and probabilistic nature of \ac{llm} offering a robust evaluation of both model performance or model compliance to certain behaviours.
However, to the best of our knowledge, no testing suite exploits the information coming out of the tests to quantify and help improve the trustworthiness of the overall software system.
Adopting \ac{ai}-based components and their integration in the development environment is still at an early stage when available.

Similarly, while we have access to tools and techniques like \emph{InTraVisTo}~\cite{DBLP:journals/corr/abs-2507-13858}, \emph{Tuned Lens}~\cite{DBLP:journals/corr/abs-2303-08112}, or \emph{LM Transparency Tool}~\cite{DBLP:journals/corr/abs-2404-07004}, which are designed specifically for the explainability of \acp{llm}, we hardly see these tools integrated in any development platform.
However, these tools can turn out to be very helpful for ``debugging a prompt'' a prompt for example, showing how the instructions give to the model affect the predicted output.

\subsection{Known Challenges}
\label{sec:known-challenges}

Recent empirical and conceptual studies highlight fundamental challenges in engineering \ac{genai}ware, and, more broadly, \ac{ai}-based sotware~\cite{schuler2026architecting}. 
Although these works address different aspects, they collectively reveal consistent patterns that point to deeper structural issues in how \ac{llm} interactions are currently designed and implemented.
A key difficulty is that prompts lack a predefined grammar, making it hard to reason about their structure and to support standard programming operations such as refactoring or reuse~\cite{10.1145/3544549.3585737}.
Recent studies highlight that testing and debugging \ac{llm}-based systems are fundamentally challenging due to non-determinism and implicit interaction logic, requiring new abstractions beyond traditional software engineering approaches~\cite{10.1145/3806396,DBLP:conf/icse-seip/NaharKBPZB25}.

Unlike traditional software components, \ac{llm}-based components exhibit inherent non-determinism, producing different outputs even under identical inputs and configurations.
As a result, correctness can no longer be assessed as a binary property, but rather as a distribution of outcomes across repeated executions, challenging established assumptions in \ac{se}, particularly in testing, verification, and reproducibility~\cite{10.1145/3806396}.
Furthermore, behavior in \ac{genai}ware is not solely defined by code but emerges from the interaction between prompts, model inference, and surrounding control logic.
Prompt construction acts as both input specification and behavioral modifier, effectively encoding parts of the system logic in natural language, yet this interaction logic is typically embedded in general-purpose code, where it remains implicit and difficult to analyze or reason about~\cite{10.1145/3806396}.

Empirical studies show that developers face persistent and diverse challenges when developing \ac{genai}ware, including prompt design, integration, reproducibility, and output interpretation.
Many of these issues are difficult to resolve in practice, with a low proportion of questions receiving accepted solutions and long response times in developer communities~\cite{DBLP:journals/tosem/ChenGCZL25}.
These findings indicate that current development practices lack adequate abstractions and methodologies to support reliable engineering of \ac{llm}-based systems.

Taken together, these findings point to a fundamental gap: the primary source of complexity in \ac{genai}ware lies not only in the models themselves, but also in the interaction logic that governs their use. 
This logic, expressed through prompt design, context management, and multi-step interactions, is typically implicit, scattered across code, and difficult to analyze, test, and maintain.
From these observations, we derive the following requirements for engineering \ac{genai}ware:
\begin{enumerate}[label=R\arabic*]
    \item \emph{Explicit interaction structure}: interaction logic should be explicitly represented rather than embedded implicitly in host language code.
    \item \emph{Separation of concerns}: deterministic control flow and probabilistic model behavior should be clearly separated.
    \item \emph{Analyzability and verifiability}: interaction structures should be amenable to validation, verification, and systematic debugging.
    \item \emph{Support for variability}: the programming model should explicitly account for non-deterministic outputs and variability across executions.
\end{enumerate}

\subsection{Positioning of this Work}
\label{sec:positioningofthiswork}

A natural question is why a new domain-specific language is needed at all.
Powerful general-purpose languages such as Python already exist, alongside a rich ecosystem of libraries for \ac{llm} orchestration, validation, and guardrails.
Indeed, many of the individual techniques discussed in this paper, like as prompt templating, structured outputs, or output validation, can be implemented already using existing frameworks.
Across multiple empirical studies, developers consistently report difficulties in prompt design, integration, and failure diagnosis, indicating that current approaches do not scale to complex workflows~\cite{10.1145/3544549.3585737}.

Our position is that the core limitation of current approaches is not expressiveness, but analyzability through an appropriate abstraction level. 
Existing approaches to building \ac{genai}ware largely rely on general-purpose programming languages and framework-based abstractions to orchestrate model interactions. 
While these approaches provide flexibility and prototyping capabilities, they typically treat prompt construction and interaction logic as ad hoc elements embedded in application code. 
As a result, they do not make interaction structure explicit, nor do they provide mechanisms for analyzing or verifying the behavior arising from such interactions.

Recent research has highlighted the need for new \ac{se} approaches that account for the non-deterministic and interaction-driven nature of LLM-based systems. 
In particular, challenges related to variability, testing, and developer experience suggest that prompt programming cannot be viewed merely as string manipulation, but requires principled abstractions to manage complexity.

In this context, we position \ac{lsl} as a \ac{dsl} designed to address the structural limitations of current practices by introducing first-class abstractions for modeling \ac{llm} interactions. 
Raher than embedding prompts and interaction logic implicitly within general-purpose code, \ac{lsl} represents interactions as explicit, structured programs.

Specifically, \ac{lsl} addresses the requirements identified in \Cref{sec:known-challenges} with the following features:
\begin{enumerate}[label=F\arabic*]
    \item \ac{lsl} provides explicit constructs for defining interaction blocks and workflows, making prompt-based logic visible and structured.
    \item \ac{lsl} separates deterministic control logic from probabilistic model execution, enabling clearer reasoning about system behavior.
    \item By enforcing structural constraints and well-defined interaction boundaries, \ac{lsl} enables static analysis and systematic validation of interaction logic.
    \item \ac{lsl} exposes interaction boundaries and constraints that make variability explicit and manageable within the programming model.
\end{enumerate}

\ac{lsl} plays a role analogous to SQL for databases or regular expressions for pattern matching: it does not supersede host languages, but provides a specialized abstraction that enables reasoning, optimization, and verification that would be impractical in unrestricted code.
Overall, \ac{lsl} shifts prompt programming from an implicit and ad hoc practice to an explicit, structured, and analyzable software engineering activity, providing a principled foundation for developing and maintaining \ac{genai}ware.

\section{LLM Scripting Language}
\label{sec:llmscriptinglanguage}

In this section, we detail our proposed \ac{dsl}, \ac{lsl}, as a way to control \acp{llm} within pipelines and workflows (\Cref{sec:framework}), we explain how we envision the approach to constrain and exploit \acp{llm} generative capabilities (\Cref{sec:beyondprompting}) and the interaction structure (\Cref{sec:beyondlinearinteractions}) in this framework.
Moreover, we provide overview of the \ac{lsl} core language model (\Cref{sec:corelanguagemodel})
Finally, we comment on the expected user base of our \ac{dsl} (\Cref{sec:stakeholders}).

\subsection{Proposed Framework}
\label{sec:framework}

\begin{figure}[!ht]
\begin{center}
    \includegraphics[width=.755\columnwidth]{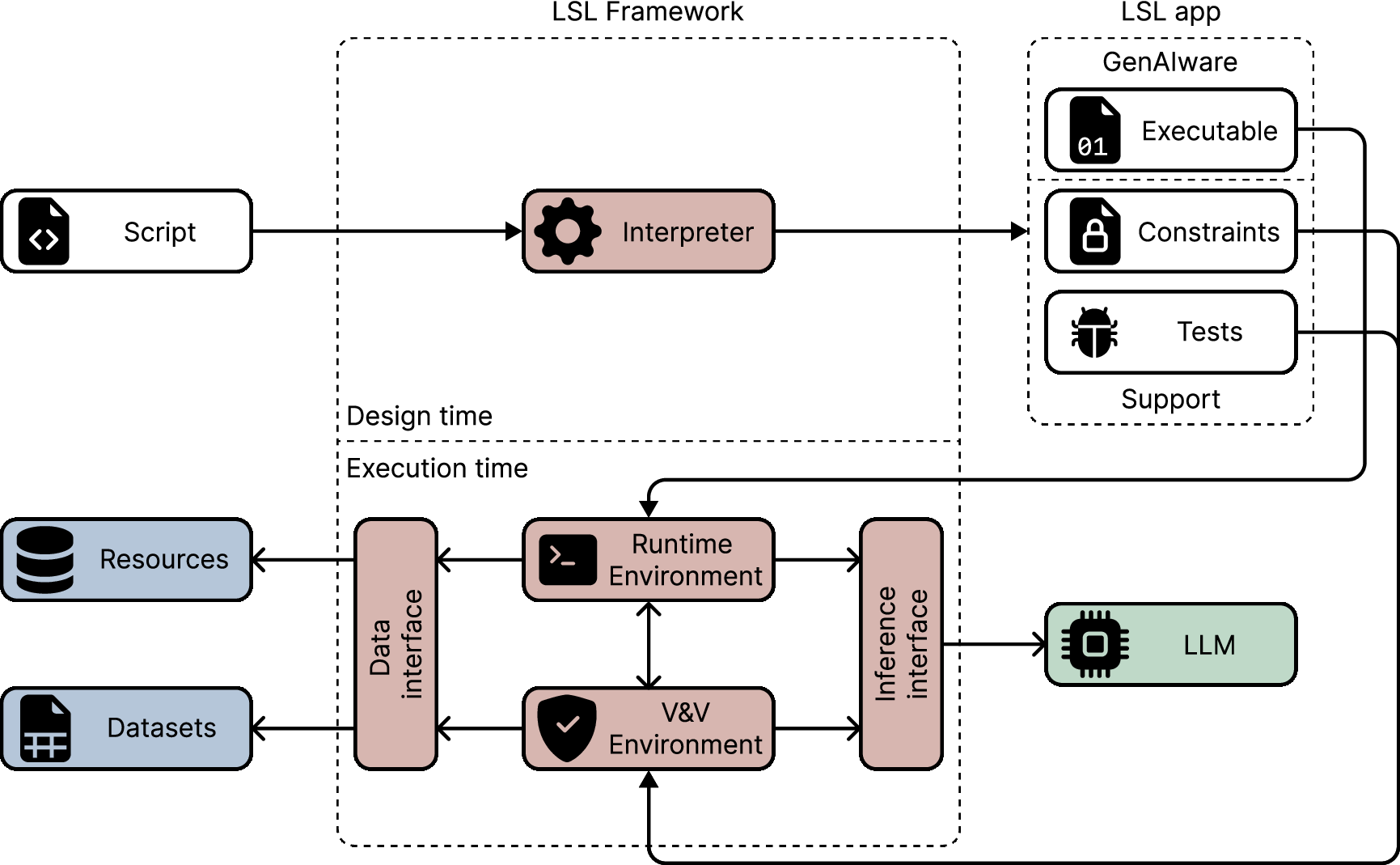}
\caption{Idea behind \ac{lsl}: the script is converted into an application bunding an executable (which is ingested by the runtime environment) and the support material composed of the constraints to satisfy and the tests to validate the program behaviour (which are ingested by the verification and validation environment). The data interface provide access to datasets (for benchmarking, performance estimate, etc.) and to resources (document bases, knowledge bases, etc.), the inference interface warps the access to the \ac{llm}.}
\label{fig:idea}
\end{center}
\end{figure}

We propose \ac{lsl}: a \ac{dsl} to format and constrain the input and output of an \ac{llm}.
The idea is to develop a scripting language, its interpreter and the surrounding development tools to improve the guarantees on many aspects concerning the use of \acp{llm} and make them more autonomous.
The interpreter running the script should interact with the \ac{llm} either communicating with it locally or remotely in a seamless way to the user.

Drawing from \ac{hci} theory, \ac{lsl} can be understood as a \emph{cognitive scaffold} that restructures how users (\ac{se} practitioners) formulate, execute, and reason about interactions with language models. 
Rather than relying on ad hoc prompt engineering, the \ac{lsl} provides explicit abstractions for control flow, data dependencies, and intermediate reasoning steps, thereby externalizing mental processes that would otherwise remain implicit. 
In line with theories of \emph{external cognition} (i.e., cognitive offloading)~\cite{10.5555/200550} and distributed cognition~\cite{DBLP:journals/tochi/HollanHK00,Obendorf2009}, the language acts as a representational medium that reduces cognitive load, improves transparency, and supports more systematic debugging and reuse. 

Prior work shows that existing prompt engineering tools provide little support for forming accurate mental models of \ac{llm} behavior, forcing users to rely on trial-and-error and intuition), further motivating structured abstractions such as \ac{lsl} as cognitive scaffolding~\cite{10.1145/3544549.3585737}.
By constraining interaction patterns and making reasoning structure visible, \ac{lsl} helps bridge the semantic gap between user intent and \ac{llm} behavior, enabling more reliable and cognitively tractable human–\ac{ai} collaboration.

The intended uses of the scripting language are
\begin{enumerate*}[label=(\roman*)]
    \item defining interaction blocks,
    \item managing the context processed by the \ac{llm},
    \item formatting input prompt to the \ac{llm};
    \item constraining output completion of the \ac{llm}.
\end{enumerate*}

We are designing this \ac{dsl} to be used by experts who need to integrate \ac{ai} in their applications or services.
The experts use \ac{lsl} to write scripts to automate some pipeline, part of an application, using \acp{llm}.
The end user, or other services, can benefit from these \ac{ai}-based applications using these pipelines. 
In \Cref{fig:idea} we provide an high-level view of the development environment.
An \ac{lsl} script already contains all the essential information to generate some form of executable (which basically is the \ac{genai}ware) together with the information to constrain the input and output and to validate the interactions automatically.
All this can be generated and packed together by the language interpreter into a \ac{lsl} application.
The runtime environment and the verification and validation environment work together during execution to run the application while enforcing and monitoring the runtime behaviour during the interactions.
Interfaces are provided both on the \ac{llm} side, to provide inference, and to the data side, to allow the use of both datasets (for testing, calibration, few-shot learning, etc.) and external resources (like document bases, knowledge graphs, etc.).

\begin{figure}[!ht]
    \centering
    \includegraphics[width=.7\columnwidth]{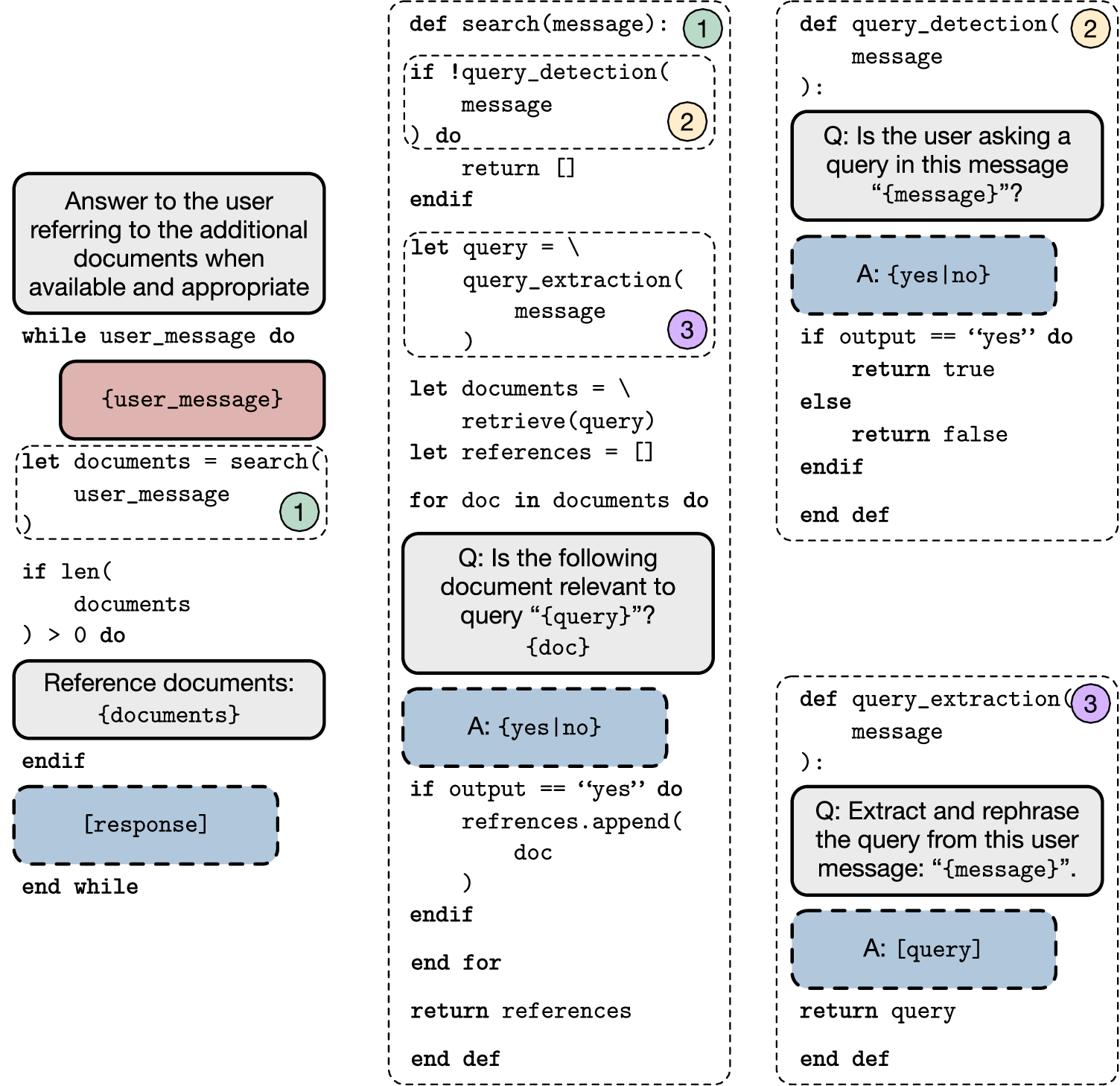}
    \caption{Example semi-structured use case (system messages, instructions, data, and subroutine outputs): knowledge-grounded chat.}
    \label{fig:ssuc}
\end{figure}
\begin{figure}[!ht]
    \centering
    \includegraphics[width=.7\columnwidth]{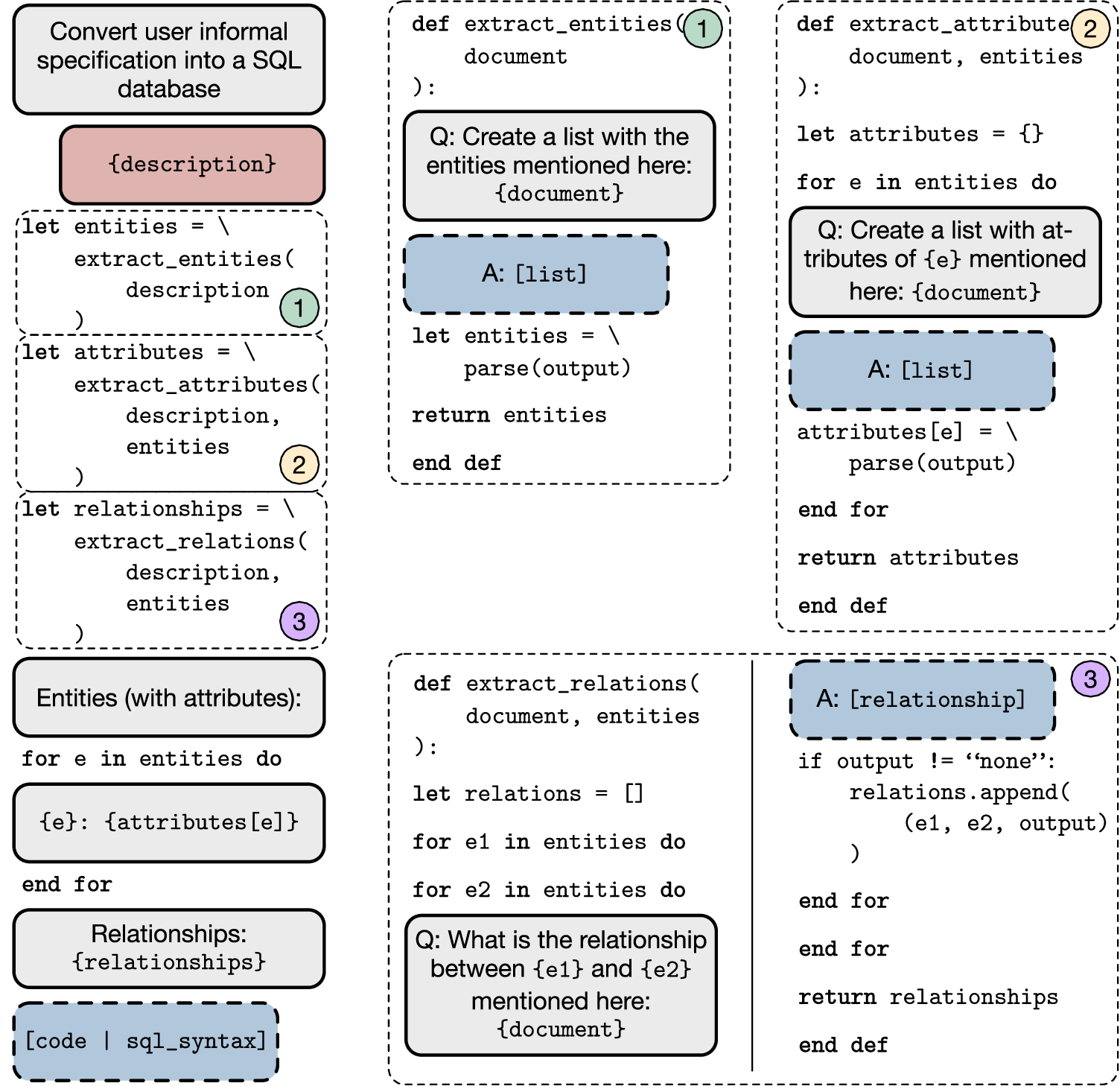}
    \caption{Example structured use case (system messages, instructions, data and subroutine outputs): database creation}
    \label{fig:suc}
\end{figure}

\ac{lsl} is thought to be used in any kind of interaction, from unstructured chit-chats to more complex structured or semi-structured use cases, like knowledge-grounded conversations, and automating the creation of databases, as we show in the examples of \Cref{fig:ssuc,fig:suc}.
Considering the semi-structured use case of the knowledge-grounded chat, as represented in \Cref{fig:ssuc}, we can see how \ac{lsl} can be used to provide a global structure for the dialogue, alternating user messages and model responses.
Before triggering the \ac{llm} completion to get the response, we can issue a call to a sub-routine that takes care of gathering relevant documents.
This sub-routine, in turn, issues two consecutive calls to as many sub-routines for query identification and extraction.
Lastly, the approach uses the \ac{llm} to further filter retrieved documents, all in a separate context.
Conversely, structured use cases find a lot of compatible applications in process automation, like in the example of database creation from informal specifications of \Cref{fig:suc}.
In this second use case, we can see how \ac{lsl} can be used to issue three consecutive calls to as many sub-routines to identify entities, attributes, and relationships, all in separate contexts.
From the extracted information, \ac{lsl} can inject another piece of the prompt similar to \ac{cot} to ease the generation of the SQL code with the definitions of the tables; moreover, with \ac{lsl}, we can constrain the last completion to stick to the SQL syntax to enforce well-formed code.

When we mention \enquote{constraining the output of the \ac{llm}}, we mean limiting the output to a subset of possible strings, whether they are enumerated or specified with some formalism, as we detail in \Cref{sec:beyondprompting}.
Instead, when we mention \enquote{controlling the interaction with the \ac{llm}} we mean properly adding or removing pieces of the context whenever prompting the \ac{llm}, as we further outline in \Cref{sec:beyondlinearinteractions}.
In \Cref{fig:ssuc,fig:suc}, we show how we plan to constrain the output:
We enforce the generated answer to some questions to be \enquote{yes} or \enquote{no}, as shown in both \Cref{fig:ssuc} and \Cref{fig:suc}, or to stick to some specific syntax, as displayed in \Cref{fig:suc}.
Similarly, we show how we plan to control the scope of the interaction to a single question and its answer rather than retaining all user input when selecting documents, like in \Cref{fig:ssuc}, or when extracting entities and their relationships, like in \Cref{fig:suc}.

\begin{table}[!ht]
\caption{Key differences between existing Python-based frameworks and the proposed \ac{lsl}}
\begin{center}
\begin{tabular}{lll}
\toprule

\textbf{Aspect} & \textbf{Python-based framework} & \textbf{\ac{lsl}} \\

\midrule

Interaction logic & Embedded in host language & First-class scripted artifact \\
Output constraints & Runtime validation & Part of execution semantics \\
Static analysis & Limited & Possible (termination, reachability) \\
Verification & External, ad hoc & Integrated into the language model \\
Intended user & Application developer & SE practitioner defining contracts \\
Primary goal & Productivity & Reliability and analyzability \\

\bottomrule

\end{tabular}
\end{center}
\label{tab:distinction}
\end{table}%

\ac{lsl} is therefore not proposed as a competitor to existing orchestration frameworks, but as a complementary abstraction that enables principled engineering analysis of \ac{llm} interactions.
We summarize the core distinctions of \ac{lsl} from existing Python-based frameworks in \Cref{tab:distinction}.

\subsection{Beyond Prompting}
\label{sec:beyondprompting}

Engineering the prompt to trigger the correct model completion is a non-trivial task~\cite{DBLP:conf/iclr/SanhWRBSACSRDBX22,DBLP:journals/jmlr/ChungHLZTFL00BW24}.
Besides phrasing the request to the \ac{llm} correctly in natural language, there are other difficulties.
One difficulty is making sure to provide all necessary information (e.g., task description, target labels, output format).
Another crucial difficulty concerns using correct techniques like few-shot learning to increase the probability of triggering the desired model's behaviour, whether it is solving the correct task or sticking to the correct output format.
Nevertheless, all these techniques cannot give any formal guarantee on the structure of the generated output.

Generative grammars can be the solution to tackle this issue.
In fact, they offer a formalism that can be exploited to constrain the decoding of the \ac{llm} output probabilities to well-formed strings according to a given specification.
These grammars can be applied freely to the generation process of an \ac{llm}.
There are simple use cases, like generating only digit tokens if computing the value of a mathematical expression.
Other use cases are more complex, like the example of generating syntactically correct SQL code for a query to be run in \Cref{fig:suc} or generating JSON strings that respond to a given JSON schema. %
As we discuss later in \Cref{sec:opportunities,sec:challenges}, this constrained decoding still does not ensure the soundness of the output, but at least gives guarantees on the fidelity to the desired structures.

With these grammars, we can specify the set of all acceptable and desirable completions for the current prompt.
The proper way to select the completion would be to pick the one that maximises the joint probability of its tokens according to the \ac{llm}.
However, this may not be feasible either because the set of valid strings for the grammar is infinite or because it is still too large to be processed in a reasonable time.
In these cases, resorting to search strategies like \emph{greedy search} or \emph{beam search} may help reduce the computational burden at the cost of introducing sub-optimal, but still syntactically valid, solutions.

The use of these grammars does not exclude the combined use with typical prompt engineering techniques that are still useful to make sure the probabilities in the output correlate with the task to solve.
Additionally, \ac{lsl} and its interpreter can be designed to help in prompt programming%
, offering tools for templating the context in a way that is more suitable to the \ac{llm} or by automating some passages like the retrieval of relevant information of examples for the few-shot learning from known and reliable sources.

\subsection{Beyond Linear Interactions}
\label{sec:beyondlinearinteractions}

Most applications that integrate chatbots based on \acp{llm} see the interaction with the user as a linear process where the user and chatbot alternate sending messages to each other.
However, the stateless nature of \acp{llm} makes them compatible with non-linear interaction, whether they are chats or other more structured processes~\cite{fornasiere-etal-2024-medical,DBLP:conf/avi/PucciPAM24}.
This means that we can integrate an \ac{llm} in a workflow characterised by
\begin{enumerate*}[label=(\roman*)]
    \item multiple, possibly independent, steps to solve the main task or its sub-tasks,
    \item isolated steps that do not require processing the entire context,
    \item states that needs to be passed from one step to another and
    \item forward and backward evolutions of the interaction.
\end{enumerate*}

\Cref{fig:suc} shows how both a semi-structured task, such as the knowledge-grounded chat of the example in \Cref{fig:ssuc}, and a structured task like the generation of SQL code to build a database of the example in \Cref{fig:suc} require multiple steps.
Moreover, some passages are iterative, some steps are self-contained, and many of them require defining and moving around variables: these are all premises for the development of a \ac{dsl}.
Nowadays there are tools and frameworks like \emph{Langchain}~\cite{docs/introduction/} to help define pipelines and automate processes or like \emph{NeMo Guardrails} to impose some constraints on the dialogue flow and content~\cite{DBLP:conf/emnlp/RebedeaDSPC23}, but
\begin{enumerate*}[label=(\roman*)]
    \item they often lack expressivity to deal with complex problems,
    \item they do not take into account possible optimisations to enhance performances,
    \item they are bound to one or a small set of domains (e.g., programming languages),
    \item they do not necessarily allow for rapid prototyping and deployment, and
    \item they do not exploit formal tools to provide some form of verification or validation.
\end{enumerate*}
With \ac{lsl}, we aim at easing the implementation of all these steps through a unified interface.

Moreover, some passages, like the relevant document selection in the search function of the example in \Cref{fig:ssuc}, or the attributes and relations extractions of the example in \Cref{fig:suc}, are independent and can be parallelised.
A proper integration of this feature in \ac{lsl} can allow automatic exploitation of the high degree of parallelism that \acp{dnn} naturally enable.
In fact, they can process samples in batches independently of each other, allowing, where applicable, a better utilisation of the computational resources.

Going further with non-linear interactions, we also consider error management.
In fact, despite the use of \ac{lsl} there still may be faults caused either by the underlying \ac{llm} or by the environment.
For these cases, we foresee the integration of tools to trace the source of the errors and apply correction actions, either by asking for user feedback, when applicable, or by prompting the \ac{llm} to provide a revised completion~\cite{DBLP:conf/nips/ShinnCGNY23}, given the previous faulty completion and the error.

\subsection{Core Language Model}
\label{sec:corelanguagemodel}

\ac{lsl} is designed for \emph{prompt programming in \ac{genai}ware systems}, where the primary abstraction is not computation over data, but structured interaction with a probabilistic model. 
In such systems, prompts, context, and output constraints jointly determine the behaviour of the underlying foundation model and must therefore be treated as first-class software artifacts~\cite{llm_se_roadmap}.

The design of \ac{lsl} is guided by three core principles:
\emph{interaction-centricity},
\emph{explicit context and constraints}, and
\emph{analyzability over expressiveness}.
\ac{lsl} programs describe \emph{interaction workflows} rather than purely algorithmic computations. 
Each program consists of a sequence of interactions between deterministic software components and a \ac{llm}, where each interaction is explicitly defined and scoped.
Moreover, \ac{lsl} treats context construction, scoping, and evolution as explicit language constructs, and similarly elevates output constraints to first-class elements. This reflects the central role of context engineering and output constraining in \ac{genai}ware development~\cite{llm_se_roadmap}.
Finally, consistent with \ac{dsl} design principles, \acl{lsl} prioritizes analyzability over unrestricted expressiveness, enabling inspection and reasoning about prompt-driven behaviour as software artifacts~\cite{DBLP:conf/sofsem/Mernik17}.

The fundamental unit of computation is the \emph{interaction block}, representing a single prompt--response exchange:

\begin{lstlisting}[
    backgroundcolor=\color{gray!10},
    basicstyle=\ttfamily\small,
    breaklines=true,
    frame=single,
    columns=fullflexible,
    keepspaces=true,
    showstringspaces=false,
    breakatwhitespace=true
]
def classify_sentiment(text: str)
    write Classify the sentiment of the given text  # Preamble
    write Q: '''\n{text}\n'''                       # Input
    read A: {positive|negative|neutral}             # Output
end def
\end{lstlisting}

This interaction encapsulates the context, instruction, and output constraint as a single analyzable unit.

Context scopes define which information is visible to the model:

\begin{lstlisting}[
    backgroundcolor=\color{gray!10},
    basicstyle=\ttfamily\small,
    breaklines=true,
    frame=single,
    columns=fullflexible,
    keepspaces=true,
    showstringspaces=false,
    breakatwhitespace=true
]
@context(sys_message, history: list[mapping[str, str]])
def chat()
    while true
        {sys_message}                                                # Preamble
        {history[-2:]}                                               # Keep only last two messages
        let message: str = read User: {user}
        append(history, {"role": "user", "content": message})
        let response: str = read Assistant: {}
        append(history, {"role": "assistant", "content": response})
        clear                                                        # Clear context
    end while
\end{lstlisting}

This prevents unintended accumulation of context and enables reasoning about information dependencies.

Outputs are constrained using formal specifications, such as enumerations, schemas, or grammars.

\begin{lstlisting}[
    backgroundcolor=\color{gray!10},
    basicstyle=\ttfamily\small,
    breaklines=true,
    frame=single,
    columns=fullflexible,
    keepspaces=true,
    showstringspaces=false,
    breakatwhitespace=true
]
import json

...

def extract_entities(document: str): json[list[str]]
    Extract entities from the given document                           # Preamble (write implicit)
    '''\n{text}\n'''                                                   # Input (write implicit)
    let entities: json[list[str]] = read Entities:\n{json[list[str]]}  # Output "```json\n...\n```"
    return entities
end def
\end{lstlisting}

More expressive constraints can be defined using grammars:

\begin{lstlisting}[
    backgroundcolor=\color{gray!10},
    basicstyle=\ttfamily\small,
    breaklines=true,
    frame=single,
    columns=fullflexible,
    keepspaces=true,
    showstringspaces=false,
    breakatwhitespace=true
]
import sql

...

def generate_sql(
    entities: list(str), attributes: mapping(str, list[str], relationships: list[tuple[str, str, str]
)
    Create a SQL database given the following entities and relationships.
    Entities
    for let e: str in entities
        {e}:
        for let a: str in attributes[e]
        \t- {a}
        end for
    end for
    for let e1: str, e2: str, r: str in relationships
        {r}({e1}, {e2})
    end for
    let code: sql[str] = read {sql[str]}
end def
\end{lstlisting}

\ac{lsl} provides constructs for sequencing, branching, and iteration.

\begin{lstlisting}[
    backgroundcolor=\color{gray!10},
    basicstyle=\ttfamily\small,
    breaklines=true,
    frame=single,
    columns=fullflexible,
    keepspaces=true,
    showstringspaces=false,
    breakatwhitespace=true
]
def search(query: str): list[str]
    ...
end def

def answer_with_docs(message: str, docs: list[str]): str
    ...
end def

def answer(message: str): str
    ...
end def

...

let response: str
if detect_query(message) do
    let docs: list[str] = search(message)
    response = answer_with_docs(message, docs)
else
    response = answer(message)
endif

...
\end{lstlisting}

Execution can be monitored and adapted dynamically.

\begin{lstlisting}[
    backgroundcolor=\color{gray!10},
    basicstyle=\ttfamily\small,
    breaklines=true,
    frame=single,
    columns=fullflexible,
    keepspaces=true,
    showstringspaces=false,
    breakatwhitespace=true
]
import python

def correct(spec: str, code: python[str], error str): python[str]
    ...
end def

...

def generate_code(spec: str, tests: list[callable[python[str]]): str
    Write a Python function with the following specifications.  # Preamble (write implicit)
    {spec}                                                      # Input (write implicit)
    let code: python[str] = read {python[str]}                  # Output
    let counter: int = 0
    try
        label retry
        for let t: callable in test:
            t(code)
        on error:
            code = correct(spec, code, error)
            if counter <= 2 do
                counter += 1
                goto retry
            end if
        end on
    end try
end def
\end{lstlisting}

In our intended execution model, an \ac{lsl} program is interpreted as a sequence of steps alternating between deterministic computation and probabilistic model invocation. 
Each interaction block is evaluated through the following stages:

\begin{enumerate}
    \item \emph{Context construction}, assembling relevant inputs and state;
    \item \emph{Prompt instantiation}, materializing the instruction template;
    \item \emph{Constrained generation}, invoking the LLM under output constraints;
    \item \emph{Validation and binding}, checking conformance and storing results.
\end{enumerate}

This execution model introduces a clear separation between conventional control flow and probabilistic behaviour, which is essential for enabling analysis and reasoning~\cite{llm_se_roadmap}.

Consider a two-step workflow:

\begin{lstlisting}[
    backgroundcolor=\color{gray!10},
    basicstyle=\ttfamily\small,
    breaklines=true,
    frame=single,
    columns=fullflexible,
    keepspaces=true,
    showstringspaces=false,
    breakatwhitespace=true
]
def classify_sentiment(text: str): [positive|negative|neutral]
    ...
end def

def summarize_issue(text: str): str
    ...
end def

...

let sentiment: [positive|negative|neutral] = classify_sentiment(review)
if sentiment == "negative" do
    let summary: str = summarize_issue(review)
    ...
end if

...
\end{lstlisting}

This corresponds to two interaction blocks evaluated sequentially, with explicit data flow between them.

\ac{lsl} semantics are defined in terms of structured interactions with explicit control over both context and output space.

Context is explicitly constructed and scoped for each interaction.

\begin{lstlisting}[
    backgroundcolor=\color{gray!10},
    basicstyle=\ttfamily\small,
    breaklines=true,
    frame=single,
    columns=fullflexible,
    keepspaces=true,
    showstringspaces=false,
    breakatwhitespace=true
]
@context(history: list[mapping[str, str])
def answer_with_docs(documents: list[str]): str
    let message str: history[-1]["content"]
    new                                                                              # Start new context
    Answer the following message referring to the given documents when appropriate.  # Preamble
    Reference documents:
    for let doc: str in documents:
        '''\n{doc}\n'''
    Q: {message}                                                                     # Input
    let response: str = read {response}                                              # Output
    return response
end def
\end{lstlisting}

This avoids interference between unrelated steps and ensures predictable behaviour.

Output validity is defined by conformance to constraints. 
\ac{lsl} enforces \emph{structural correctness} by restricting the set of admissible outputs.

\begin{lstlisting}[
    backgroundcolor=\color{gray!10},
    basicstyle=\ttfamily\small,
    breaklines=true,
    frame=single,
    columns=fullflexible,
    keepspaces=true,
    showstringspaces=false,
    breakatwhitespace=true
]
def detect_query(message): bool
    Q: Is the user asking a query in this message: "{message}"?
    let response: [yes|no] = A: {yes|no}
    if response == "yes" do
        return true
    else
        return false
    endif
end def
\end{lstlisting}

The model is constrained to produce one of the allowed values, ensuring compatibility with downstream control flow.

Our objective with \ac{lsl} is to enable reasoning about:
\begin{enumerate*}[label=(\roman*)]
    \item control-flow properties (e.g., reachability, termination),
    \item structural validity of outputs,
    \item consistency of interaction contracts,
    \item presence of recovery strategies.
\end{enumerate*}

For example, given a grammar-constrained output:

\begin{lstlisting}[
    backgroundcolor=\color{gray!10},
    basicstyle=\ttfamily\small,
    breaklines=true,
    frame=single,
    columns=fullflexible,
    keepspaces=true,
    showstringspaces=false,
    breakatwhitespace=true
]
import json
import service_xyz

...

def call_api(intent: str, params: json[mapping[str, str]])
    Generate the API call for the given intent and parameters.  # Preamble (write implicit)
    Intent: {intent}                                            # Input (write implicit)
    Parameters: {params}
    let call: service_xyz[json] = read {service_xyz[json]}      # Output
    call()
end def
\end{lstlisting}

one can ensure that all generated API calls conform to the expected syntax.

However, \ac{lsl} does not guarantee:
\begin{enumerate*}[label=(\roman*)]
    \item factual correctness,
    \item absence of hallucinations,
    \item semantic validity of outputs.
\end{enumerate*}

These limitations arise from the probabilistic nature of LLMs, whose outputs are generated based on learned distributions rather than deterministic computation.

\subsection{Stakeholders}
\label{sec:stakeholders}

\ac{lsl} is designed primarily for software engineers who need to integrate \ac{llm}-powered functionality into applications and services, but who may lack deep expertise in machine learning or natural language processing.
These practitioners understand their application domains and requirements well, but face barriers when attempting to prompt \acp{llm} effectively: they must navigate complex machine learning or deep learning frameworks, manage probabilistic outputs, and ensure reliability in production systems.
By providing a scripting interface that abstracts away low-level \ac{llm} implementation details while still offering fine-grained control over interactions, \ac{lsl} gives these professionals the opportunity to focus on solving domain-specific problems rather than
\begin{enumerate*}[label=(\roman*)]
    \item dealing with the machine learning infrastructure, and
    \item managing the several tools necessary to build the intended component.
\end{enumerate*}
Secondary stakeholders include DevOps engineers responsible for deploying and maintaining \ac{ai}-powered services, as well as quality assurance professionals who need to verify and validate \ac{llm}-based components within larger software systems.

We assume that \ac{lsl} users possess core \ac{se} competencies rather than specialized \ac{ai} expertise.
Specifically, users should be proficient in programming under imperative or procedural paradigms (e.g., using languages like C, Bash, or Python) and comfortable with standard programming constructs such as control flow, functions, and error handling.
They should have practical knowledge of \ac{se} practices, including testing, debugging, and verification techniques, as well as basic familiarity with DevOps workflows like \emph{continuous integration} or \emph{continuous delivery} pipelines.
While users need not be machine learning researchers, they should understand \acp{llm} at a conceptual level; this includes \acp{llm} capabilities, limitations (e.g., hallucinations, probabilistic outputs), as well as basic prompt engineering techniques such as few-shot learning.
Experience with \acp{dsl} (e.g., regular expressions, build scripts) and APIs is beneficial, as is the ability to design multi-step workflows and data pipelines.
Finally, users should be willing to engage with lightweight formalism, such as defining constraints through JSON schemas or grammar-like specifications, without requiring expertise in formal verification methods.

\section{Illustrative Example}
\label{sec:illustrative-example}

To illustrate the implications of the proposed \ac{lsl} programming model, we analyse a representative application example: the generation of SQL code from informal user specifications, as introduced in \Cref{fig:suc}. 
This scenario is representative of \ac{genai}ware systems where \acp{llm} are used to produce structured artifacts that are subsequently consumed by deterministic software components.

The SQL generation pipeline in \Cref{fig:suc} involves multiple steps, including:
\begin{enumerate*}[label=(\roman*)]
    \item entity extraction,
    \item attribute identification,
    \item relationship extraction, and
    \item SQL code generation.
\end{enumerate*}
Each step relies on \ac{llm} outputs that must conform to expected structures to ensure correct downstream processing.

We consider two alternative implementations of the same workflow.
The former is a \emph{Python-based implementation}, combining a general-purpose language with existing frameworks (e.g., LangChain for orchestration and Pydantic for output validation).
The latter is an \emph{\ac{lsl}-based implementation}, where interactions, context management, and output constraints are expressed as first-class constructs in a dedicated prompt-programming language.

Rather than evaluating model accuracy, which depends on the underlying \ac{llm}, we focus on \emph{engineering-relevant failure modes} that affect the reliability and maintainability of \ac{genai}ware systems. 
In particular, we consider:

\begin{itemize}
    \item \emph{Invalid output rate}: the frequency with which generated outputs do not conform to the expected structure (e.g., malformed SQL).
    \item \emph{Pipeline breakage}: the extent to which invalid or unexpected outputs propagate and disrupt downstream steps.
    \item \emph{Recovery effort}: the effort required to detect, localize, and correct failures during execution.
\end{itemize}

\paragraph{Analysis}

The SQL generation scenario exposes several concrete failure cases:
\begin{enumerate*}[label=(\roman*)]
    \item \emph{Malformed SQL syntax}, e.g., missing parentheses or invalid keywords, causing execution errors in downstream systems;
    \item \emph{Inconsistent intermediate representations}, e.g., entities extracted without corresponding attributes, leading to incomplete schema definitions;
    \item \emph{Missing or incorrect relationships}, resulting in logically inconsistent database schemas;
    \item \emph{Incorrect control decisions}, e.g., failure to detect whether a user query requires structured processing, leading to incorrect pipeline execution.
\end{enumerate*}

In Python-based implementations, these failures are managed in an ad hoc manner. 
Moreover, output validation is typically performed post hoc and control-flow decisions depend on possibly invalid or missing outputs. 
For instance, malformed SQL is typically detected only at execution time, while inconsistencies between entities and attributes may propagate across multiple steps before being detected. 
Validation logic (e.g., JSON schema checks or type validation) is applied after generation, and control flow decisions depend on outputs that may be missing, malformed, or semantically incorrect. 
Often these failures are caught by some mechanism that feeds back the error to \ac{llm} in an attempt to fix it, leading to the so called ``Code Roulette''~\cite{prompting-effecting-variability}.

In contrast, \ac{lsl} externalizes these concerns into the interaction model. 
Intermediate steps such as entity extraction and relationship detection are associated with explicit output constraints (e.g., lists, schemas), while final SQL generation can be constrained using formal grammars. 
As a result, structural violations can be detected or prevented through constrained decoding at the boundary of each interaction block, reducing the likelihood of error propagation.

Even in the absence of formal syntax, prompts exhibit implicit semantic structure (e.g., preambles, examples, templates), which can be inferred and operationalized~\cite{10.1145/3544549.3585737}.
\ac{lsl} builds on this intuition by making such structure explicit and enforceable.

\begin{table}[!ht]
\caption{Comparison of failure modes in Python-based frameworks and LSL}
\begin{center}
\begin{tabular}{lll}
\toprule

\textbf{Scenario} & \textbf{Python-based framework} & \textbf{\ac{lsl}} \\

\midrule

Malformed output & Runtime exception or invalid data propagation & \makecell[l]{Structurally prevented\\(or detected at interaction boundary)} \\

Missing function call & Silent failure or incorrect branching & Enforced by interaction specification \\

Error localization & Distributed across program logic & Localized to interaction block \\

Recovery handling & Ad hoc, application-specific & Explicit, integrated in execution model \\

\bottomrule

\end{tabular}
\end{center}
\label{tab:eval}
\end{table}

\Cref{tab:eval} summarizes key differences between the two approaches.

\paragraph{Discussion and Limitations}

This application example shows that \ac{lsl} improves robustness in structured generation scenarios such as SQL synthesis. 
By constraining outputs and isolating interactions, \ac{lsl} reduces both the likelihood and the impact of structural failures. 
In particular, grammar-based constraints are effective in ensuring syntactically valid SQL, while explicit interaction boundaries prevent the propagation of invalid intermediate representations. 
Moreover, \ac{lsl} improves interpretability by making interaction logic explicit.

However, improvements introduced by \ac{lsl} come with trade-offs 
In fact, the improvements we reported in this example scenario are limited to \emph{structural correctness}. 
\ac{lsl} does not guarantee that the generated SQL is semantically correct with respect to the user's intent, nor that the extracted entities and relationships reflect the true meaning of the input specification.
Moreover, introducing a \ac{dsl} requires developers to learn new abstractions and integrate an additional execution layer.
Additionally, \ac{lsl} addresses structural and interaction-level failures but does not eliminate model-level issues such as hallucinations or ambiguity in natural language instructions.

Overall, the application example suggests that \ac{lsl} shifts failure modes from \emph{implicit and distributed} (in Python-based implementations) to \emph{explicit and localized} (in prompt programs), thereby improving analyzability and maintainability without altering the underlying probabilistic nature of \acp{llm}.

\section{Opportunities}
\label{sec:opportunities}

In this section, we report our considerations about the opportunities given by \ac{lsl} for reliability (\Cref{sec:reliability}), robustness (\Cref{sec:robustness}), and trustworthiness (\Cref{sec:trustworthiness}).

\subsection{Reliability}
\label{sec:reliability}

Concerning the reliability, which we define as the ability of the \ac{llm} to \enquote{perform as required, without failure}~\cite{ai2023artificial}, we focus on the use cases where the \ac{llm} has to generate a precise output, like an exact string out of $k$ known strings (e.g., in classification), or, more importantly, it has to deal with the external environment (e.g., calling a function from an API or writing some code).
Often, functionalities to interact with the environment are implemented with \emph{function calling}: the \ac{llm} generates a response matching a specific pattern that triggers the function call.
The information about the functions is provided to the \ac{llm} in the context as part of the task instructions.

Classification problems require that the \ac{llm} generates the correct name of the class, and the function calling approach requires that the model
\begin{enumerate*}[label=(\arabic*)]
    \item would generate the call,
    \item that it would generate it when required, and
    \item that it would generate it correctly.
\end{enumerate*}
However, despite the fine-tuning of the few-shot learning on samples for classification or requiring function calls, we cannot ensure that the output will comply with the desired one.
As a result, a system that relies on the \ac{llm} taking care of, for example, calling a function to complete autonomously some workflow would lose reliability.

Conversely, using \ac{lsl} can ensure that, for example, if a function call is needed, the scripted parts of the interaction will trigger that call.
Moreover, the possibility to constrain the output will ensure that the parameters passed to the function call are syntactically correct.
Similarly, if we need the \ac{llm} to generate some code at runtime (e.g., run some bash command to check the status of a server), we can ensure that it is at least syntactically correct and that the script is executed.
As a result, we would have more autonomous \acp{llm} running inside more reliable systems.

Despite the problem staying, the \ac{llm} introduces errors (as we further detail in \Cref{sec:remainingthreats}).
We can ensure that everything the \ac{llm} generates will comply with the required outputs to the following steps by constraining the output to follow specific patterns.
This can at least ensure that one of the classes to choose from is actually selected or that a function is called whenever it is needed without failing because of, e.g., missing a \texttt{`;'}.
This avoids breaking down the entire pipeline and makes it more reliable.

\begin{tcolorbox}[
    title=Reliability, %
    left=1mm,
    right=1mm,
    top=1mm,
    bottom=1mm,
    fonttitle=\bfseries, %
    boxrule=0.5pt, %
    arc=5pt, %
    width=\columnwidth %
]
We plan to tackle the reliability of generated content by constraining the decoding to obey a given structure or to target only a known, possibly finite, set of outputs.
\end{tcolorbox}

\subsection{Robustness}
\label{sec:robustness}

\acp{llm} are \acp{dnn} and as such they are designed to be robust, where robustness is intended as the ability to avoid spurious correlations and patterns~\cite{DBLP:books/daglib/0040158,DBLP:conf/nips/BrownMRSKDNSSAA20}.
Nevertheless, these models are not error-proof, and often, ambiguous or poorly formulated inputs can cause them to fail.
The same applies to use cases where there is the need for some multi-step reasoning or some additional or more updated information is needed.

A first measure to improve robustness in these cases is to automate the use of the techniques we presented in \Cref{sec:capabilities}.
For example, given a known classification task where there are some reference data sets available, \ac{lsl} can offer support for automatically retrieving proper prompts and examples for few-shot learning or \ac{cot} reasoning.
In the case of few-shot learning, it can also automatically tune the number of shots.
Similarly, \ac{lsl} can offer tools to automatically search trusted documents and knowledge bases to add useful information to the context and help the \ac{llm} generate the correct response.
All these features are thought to improve the robustness by automating the use of techniques designed to help the accuracy of \acp{llm}.

A complementary robustness measure, especially in the interaction with the external environment, can include error management.
We can fall back to a semi-automatic process by introducing a human in the loop in case of failures.
Otherwise, we can have the \ac{llm} try to fix the issue itself.
If we can trace back to the point where the error was created, we can prompt the \ac{llm} to revise its response in light of the error message.
In this sense, there already exist frameworks to use textual feedback (i.e., the error message) to have the \ac{llm} correct itself~\cite{DBLP:conf/nips/ShinnCGNY23}.
As before, integrating these solutions in \ac{lsl} will not ensure the absence of errors, but it can improve the robustness of a system using the \ac{llm}.

\begin{tcolorbox}[
    title=Robustness, %
    left=1mm,
    right=1mm,
    top=1mm,
    bottom=1mm,
    fonttitle=\bfseries, %
    boxrule=0.5pt, %
    arc=5pt, %
    width=\columnwidth %
]
We plan to tackle robustness in the use of \acp{llm}
\begin{enumerate*}[label=(\roman*)]
    \item by automating context management for few-shot learning, \ac{rag}, and \ac{cot} reasoning and
    \item by introducing error-handling subroutines (either with self-correction or with user feedback).
\end{enumerate*}
\end{tcolorbox}

\subsection{Trustworthiness}
\label{sec:trustworthiness}

In the following, we discuss the trustworthiness, defined as the ability of a \ac{llm} to \enquote{reflect characteristics such as truthfulness, safety, fairness, robustness, privacy, machine ethics, transparency, and accountability}~\cite{DBLP:conf/icml/Huang0WWZLGHLZL24} of applications interfacing with \acp{llm} using \ac{lsl} in terms of verification and validation (\Cref{sec:verificationandvalidation}) and in terms of explainability (\Cref{sec:explainability}).

\subsubsection{Verification and Validation}
\label{sec:verificationandvalidation}

Introducing a structured overlay to wrap the access to the \ac{llm} with \ac{lsl} opens the chance of using formal methods to verify and validate the scripted pipelines via model-checking tools.
In fact, while we cannot have guarantees that the model will not make errors in the predicted output, the constraints on the interaction structure (e.g., branches in the workflow) and on the generated output (e.g., code) introduced by \ac{lsl} are compatible with this kind of analysis.
Using formal tools for verification and tools for validation allows us to add guarantees on the constraints imposed to the \ac{llm} and on the outcome of the scripts realised with \ac{lsl}~\cite{ai2023artificial}.
For instance, one can verify that the interaction scripted with \ac{lsl} always terminates or that there are no unreachable or dead nodes in the computation graph.
Looking at more concrete use cases, we can also verify the generated JSON to call a remote API always matches the API constraints by looking at the grammar constraining the \ac{llm} output.

Moreover, we can take into account the probabilistic nature of the \ac{llm} and, where applicable, integrate testing methodologies based on
\begin{enumerate*}[label=(\arabic*)]
    \item fuzzying, adversarial, or evolutionary testing~\cite{DBLP:journals/corr/abs-2503-00481,DBLP:journals/tse/CorboBDLSC25} to expand the coverage automatically, and
    \item statistical test, to provide a confidence bound on the results of the tests~\cite{DBLP:journals/corr/abs-2503-00481}.
\end{enumerate*}
The latter would additionally allow the use of results on the model's accuracy on a specific task or sub-task and apply probabilistic model checking to measure the likelihood of certain behaviours or outcomes.
For example, if we need to apply sentiment analysis on a piece of text as part of a pipeline and we have a reference data set of documents, possibly coming from the same domain, we can use it to test the model, get the likelihood of committing a classification error, and use this information to verify properties of our system.
Additionally, given insights on model calibration (i.e., how well the predicted probabilities match the model's confidence), we can perform runtime analysis for a specific input, extending the set of guarantees.

\subsubsection{Explainability}
\label{sec:explainability}

Explainability with \acp{dnn}, in general, and \acp{llm}, in particular, poses an important requirement to make \ac{ai}-based application more \emph{transparent}~\cite{DBLP:conf/icml/Huang0WWZLGHLZL24}.
Besides transparency, which helps users understand the application's behavior, explainability can also serve as a debugging tool by highlighting which parts of the prompt caused faulty output and allowing users to correct such parts, for example, by rephrasing ambiguous instructions.  %

For the scope of \ac{lsl}, we see explainability adopted in two different ways.
The former solution is to plug in external frameworks, thus re-using pre-existing tools and approaches~\cite{DBLP:journals/tist/ZhaoCYLDCWYD24}.
The latter solution, more specific to \ac{lsl}, is to use the structure in the interaction imposed by the script to trace back the outcome.
The two proposed solutions are not mutually exclusive and can be combined to get better insights of what led to a specific result or prediction.

\begin{tcolorbox}[
    title=Trustworthiness, %
    left=1mm,
    right=1mm,
    top=1mm,
    bottom=1mm,
    fonttitle=\bfseries, %
    boxrule=0.5pt, %
    arc=5pt, %
    width=\columnwidth %
]
We plan to tackle trustworthiness in terms of
\begin{enumerate*}[label=(\roman*)]
    \item verification and validation, offering support for verification of the script code and probabilistic checking of the procedure described by the script, and
    \item explainability, offering support for internal solutions based on the script structure and external explainability tools.
\end{enumerate*}

\end{tcolorbox}

\section{Challenges}
\label{sec:challenges}

In this section, we report the challenges that we foresee to develop \ac{lsl} in terms of technical difficulties (\Cref{sec:technicalaspects}), standardisation of the interfaces (\Cref{sec:standardisation}), threats related to the use of \acp{llm} (\Cref{sec:remainingthreats}), and the ethical concerns (\Cref{sec:ethicalconcerns}).

\subsection{Technical Barrier}
\label{sec:technicalaspects}

Developing a scripting language like \ac{lsl} is complex, encompassing multiple \ac{se} areas including language and APIs design, concurrency, and performance~\cite{nystrom2021crafting}.

Designing the language with the appropriate level of abstraction is critical to ensure the right balance between expressivity, abstraction, and usability.
The syntax should allow the integration of script code, input templating, and output constraining in a seamless way from the developer's point of view, easing the management of context and the interaction with the model.
Particularly, output constraints, which are critical to reliability, should be analysed carefully to allow specifying both simple constraints, like alternative completions, to more complex constraints, like the syntax of a programming language.

Similarly, designing the APIs to simplify the access to functionalities like likelihood prediction, generation, embedding, or to access other resources is fundamental to offer a complete suite to the developers.
Thus, we need to understand which are the essential operations required to interact with the \ac{llm} and the environment.
This is strictly connected to the standardisation challenge we introduce in \Cref{sec:standardisation}, as providing a unified interface to models and resources is essential to effectively communicate with all the components.

Regarding concurrency and, more generally, performance, given the computational demand of running an \ac{llm}, being able to automatically parallelize the passages in the pipeline and pack those requests that can be run concurrently in batches is essential.
The concurrency aspects are not limited to a single process, but also to concurrent processes characterised by different sessions and contexts, requiring careful management of which contexts are forwarded to the \ac{llm} and when to ensure optimal resource utilisation.
Another aspect to consider is the causal nature of \acp{llm}, which enables the exploitation of caching mechanisms by reusing partially pre-processed sequences.  %

There are also aspects related to error handling, testing, verification, and validation.
Integrating explainability with error management and testing can help track down and solve issues by understanding whether they are related to the script code, the \ac{llm}, or both.
Similarly, designing the tools to apply (statistical) model checking to verify and validate the script code at compile time and run-time poses a non-trivial task, even if we want to resort to external tools.

\begin{tcolorbox}[
    title=Technical barrier, %
    left=1mm,
    right=1mm,
    top=1mm,
    bottom=1mm,
    fonttitle=\bfseries, %
    boxrule=0.5pt, %
    arc=5pt, %
    width=\columnwidth %
]
We expect the main technical challenges to come from
\begin{enumerate*}[label=(\roman*)]
    \item the design of \ac{lsl} to be sufficiently abstract and usable plus
    \item the implementation of the interpreter for \ac{lsl} when addressing automatic \ac{llm} context management and resource usage optimisation.
\end{enumerate*}
\end{tcolorbox}

\subsection{Standardization}
\label{sec:standardisation}

Standardizing the interfaces is crucial for several reasons, including \emph{interoperability} of different modules, \emph{consistency and predictability} of behaviours, and \emph{ease of integration}.
In this context, we are interested in standardising two main interfaces:
\begin{enumerate*}[label=(\arabic*)]
    \item the interface to the \ac{llm}, obviously, and
    \item the interface to external resources, like data sets or documents and knowledge bases.
\end{enumerate*}
The main challenge in these cases is either finding a good standard compatible with the most adopted solution or enforcing the adoption of a new one.

From an \ac{llm} perspective, the most adopted interface is the one provided by \emph{HuggingFace} with the \emph{Transformer} library~\cite{DBLP:conf/emnlp/WolfDSCDMCRLFDS20}.
Yet, there are other valid alternatives to self-host models, like llama.cpp~\cite{llama.cpp}, or to access them remotely via commercial APIs, like \emph{watsonx.ai}~\cite{https://www.ibm.com/products/watsonx-ai}.
Moreover, larger commercial models like GPT~\cite{https://platform.openai.com/docs/api-reference/} or Gemini~\cite{https://ai.google.dev/} have their own separate interfaces.
Large benchmark frameworks like \emph{Big Bench} and \emph{Language Model Evaluation Harness
}~\cite{lintang_sutawika_2024_14506035} tried to propose wrappers to unify the interface to \acp{llm}, yet not all APIs or inference servers support the necessary functionalities.
In fact, it is crucial to have access to a scoring method to get the likelihoods predicted by the model to enforce decoding strategies (e.g., with a grammar) independently of the \ac{llm} implementation.

Standardising the interface to external resources is crucial to automate many tasks.
In fact, having standard interfaces to data sets can help automate the retrieval of samples for few-shot learning or templates for \ac{rag}.
At the same time, having standard interfaces to document on knowledge bases can help with
\begin{enumerate*}[label=(\roman*)]
  \item making content generated by \acp{llm} more factual and based on evidence and
  \item using \acp{llm} for semantic ranking and re-ranking of search results.
\end{enumerate*}
Given standardised templates for the tasks to which \acp{llm} can be applied to, this can enable the automation of hyper-parameter tuning (e.g., number of examples in few-shot learning) and the computation of performances for later verification and validation (see \Cref{sec:verificationandvalidation}).
To our knowledge, there are resources with a large and diverse pool of datasets covering different domains and tasks, but none provide a standard that allows for automated evaluation.
Moreover, the introduction of these external sources leads to the problem of identifying trusted sources, which we discuss in \Cref{sec:ethicalconcerns}.

\begin{tcolorbox}[
    title=Standardisation, %
    left=1mm,
    right=1mm,
    top=1mm,
    bottom=1mm,
    fonttitle=\bfseries, %
    boxrule=0.5pt, %
    arc=5pt, %
    width=\columnwidth %
]
We expect standardisation challenges to be mainly related to the interfaces
\begin{enumerate*}[label=(\roman*)]
    \item towards the \ac{llm} due to existing frameworks and
    \item to external resources, like data sets and document collections.
\end{enumerate*}
\end{tcolorbox}

\subsection{Inherent Threats}
\label{sec:remainingthreats}

Throughout this paper, we have highlighted that \ac{lsl} cannot eliminate all threats and risks associated with the use of \acp{llm}~\cite{ai2023artificial}.
Nevertheless, some issues can be addressed (e.g., it is possible to enforce an output template to ensure the generated content complies with a standard), and the likelihood of others occurring can be reduced (e.g., using few-shot learning to improve classification accuracy).

Identifying and quantifying all the threats that cannot be neutralised poses a significant challenge.
In fact, these threats depend on multiple factors, including the specific \ac{llm} being used and the specific tasks and subtasks that compose pipelines relying on that model.

Undoubtedly, the problem of hallucinations, either in terms of factuality or in terms of faithfulness, poses the biggest risk.
The possibility of the \ac{llm} for generating falsehoods or not sticking to some request makes the overall system seem unreliable.
However, research in the context of hallucination detection and mitigation is advancing, although there is no definitive solution to the problem yet~\cite{DBLP:journals/csur/JiLFYSXIBMF23,DBLP:journals/corr/huangSurvey}.

\begin{tcolorbox}[
    title=Inherent threats, %
    left=1mm,
    right=1mm,
    top=1mm,
    bottom=1mm,
    fonttitle=\bfseries, %
    boxrule=0.5pt, %
    arc=5pt, %
    width=\columnwidth %
]
We expect the main challenges in dealing with long-running threats related to the use of \acp{llm}, namely erroneous predictions and hallucinations, to arise from the difficulty in identifying and quantifying the risk of their occurrence.
\end{tcolorbox}

\subsection{Ethical Concerns}
\label{sec:ethicalconcerns}

Introducing \ac{ai} in any application opens several ethical issues.
\acp{llm} in particular introduce a well-defined set of ethical and social problems one should account for when using this technology~\cite{DBLP:journals/corr/abs-2112-04359}.

Technologies aimed at automating processes using \acp{llm} often introduce problems related to \emph{job replacement} (i.e., when human tasks are delegated to some piece of software) and \emph{deskilling} (i.e., when humans lose their expertise on some tasks by stopping doing that task)~\cite{DBLP:journals/ijhci/SisonDGG24}.
The intention of \ac{lsl} is to automate processes using \acp{llm}, but instead of replacing human experts, the intention is to use their expertise to help automate such processes leveraging their domain knowledge and experience.
Moreover, human feedback and supervision are always crucial to ensure the proper functioning of the system.
However, it is up to the developer of the specific application to give support to these aspects.
\ac{lsl} and its ecosystem, through scripting interactions with \acp{llm}, can provide resources to easily integrate feedback and supervision components.

Additionally, content generated by \acp{llm} exposes to risks related to \emph{bias} and \emph{toxicity}~\cite{10.1162/coli_a_00524}, besides \emph{misinformation} and \emph{hallucination}.
These issues pose a threat towards the user who would be exposed to such content.
While there is no proper way to ensure that such content is never generated, there are solutions to
\begin{enumerate*}[label=(\roman*)]
  \item test the predisposition of \acp{llm} to generate such content~\cite{DBLP:journals/tse/CorboBDLSC25} and measure quantitatively this predisposition and
  \item to recognise this kind of harmful content, preventing sharing it with the user.
\end{enumerate*}
These tools should be integrated into the \ac{lsl} ecosystem to quantify and mitigate these risks.

\begin{tcolorbox}[
    title=Ethical concerns, %
    left=1mm,
    right=1mm,
    top=1mm,
    bottom=1mm,
    fonttitle=\bfseries, %
    boxrule=0.5pt, %
    arc=5pt, %
    width=\columnwidth %
]
We expect the main challenges related to ethical concerns to arise from problems
\begin{enumerate*}[label=(\roman*)]
    \item of replacement and deskilling,
    \item of bias and toxicity and
    \item of misinformation and hallucinations.
\end{enumerate*}
\end{tcolorbox}

\section{Evaluation}
\label{sec:evaluation}

In this section, we describe the setup of our qualitative evaluation (\Cref{sec:design}), the results of the evaluation (\Cref{sec:results}), and the mitigations for the possible threats to the validity (\Cref{sec:threats-mitigation}).

\subsection{Design}
\label{sec:design}

To investigate the perceived need for the abstractions introduced by \ac{lsl}, we conducted a survey-based study, a commonly adopted empirical method for capturing practitioners’ opinions, experiences, and expectations through structured questionnaires. 
Following established guidelines for survey-based research in \ac{se}, the study was designed as a cross-sectional survey, aiming to collect a snapshot of current practices and perceptions among practitioners working with \ac{genai}ware~\cite{DBLP:conf/acsos/CalinescuMPW20,DBLP:books/sp/08/KitchenhamP08}.

The overall objective of our survey-based study was to assess the relevance and usefulness of the features supported by \ac{lsl} from the perspective of practitioners. 
In line with this goal, the survey was structured to capture three complementary aspects: current development practices for \ac{genai}ware, the challenges encountered in practice, and the perceived usefulness and potential adoption of structured approaches to defining \ac{llm} interactions. 
This design is consistent with the use of survey-based studies to describe existing conditions and to assess attitudes and perceived needs within a target population~\cite{DBLP:books/sp/08/KitchenhamP08}.

The target population of the study consisted of practitioners and researchers with experience in developing or experimenting with \ac{genai}ware. 
Participants were recruited through convenience-based channels, including professional and academic networks, which is a common approach in exploratory surveys where the target population is difficult to enumerate precisely. 
While this approach does not support strict statistical generalization, it enabled the collection of informed opinions from possible stakeholders with relevant experience, which is appropriate given the exploratory nature of the study~\cite{DBLP:books/sp/08/KitchenhamP08}.

The survey instrument was designed iteratively to ensure clarity and relevance of the questions. 
The questionnaire includes closed questions, primarily based on Likert-scale responses. 
These questions were used to capture consistent assessments of perceived difficulty and usefulness across key dimensions. 
The design of the questions aimed to avoid ambiguity and bias, using clear and neutral wording and focusing on a single concept per item, as recommended in survey design guidelines~\cite{DBLP:conf/acsos/CalinescuMPW20,DBLP:books/sp/08/KitchenhamP08}.

The questionnaire was organized to follow a logical progression. 
It first collected background information about participants’ roles and experience with \ac{genai}ware, ensuring that respondents were in a position to meaningfully answer the subsequent questions. 
It then investigated current engineering practices and common pain points, providing a contextual grounding for the evaluation of \ac{lsl}. 
Finally, the survey introduced the concept of a structured language for defining \ac{llm} interactions and elicited participants’ perceptions regarding its usefulness, benefits, and potential limitations. 
This progression is intended to reduce bias by avoiding early priming effects and by ensuring that participants reflect on their own experience before evaluating the proposed approach.

Data collection was performed through a self-administered online questionnaire, enabling participants to complete the survey asynchronously. 
The questionnaire was designed to be concise, requiring only a few minutes to complete, in order to reduce respondent burden and improve response quality, in line with best practices for survey-based research. 
Participation was voluntary and anonymous, and no incentives were provided. \cite{DBLP:books/sp/08/KitchenhamP08}
The analysis of the collected data was performed looking at descriptive statistics of the closed questions. 

Overall, the study design prioritized the collection of practitioner perspectives and perceived needs rather than objective performance measurements. 
As such, it provided qualitative and perception-based evidence regarding the relevance of the \ac{lsl} approach, complementing the illustrative examples presented earlier in the paper.

\subsection{Results}
\label{sec:results}

The survey received $30$ responses, two were discarded as the respondent has never worked with \acp{llm} is software systems, $n = 28$ valid responses were retained for analysis.
Hereafter we report descriptive results for each section of the questionnaire, with figures organised according to the survey sections.

\begin{figure}
    \centering
    \subfloat[Question 1.\label{fig:q1}]{\includegraphics[width=.48\columnwidth]{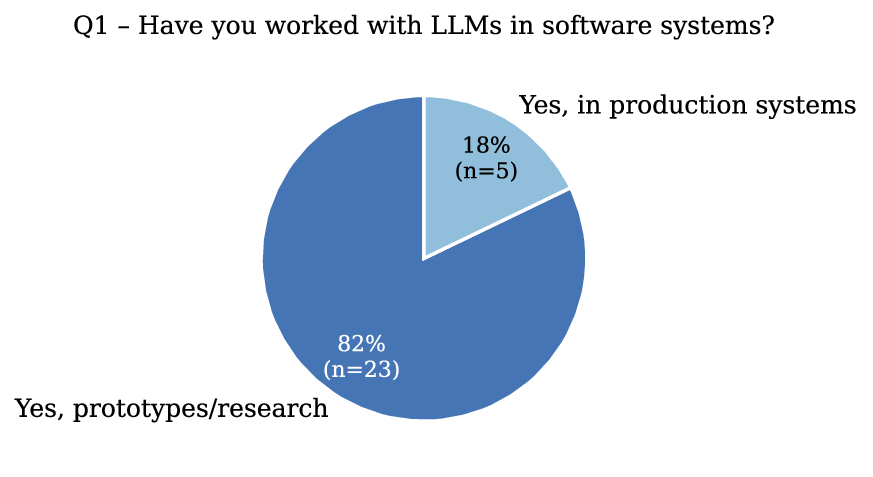}} \hfill
    \subfloat[Question 2.\label{fig:q2}]{\includegraphics[width=.48\columnwidth]{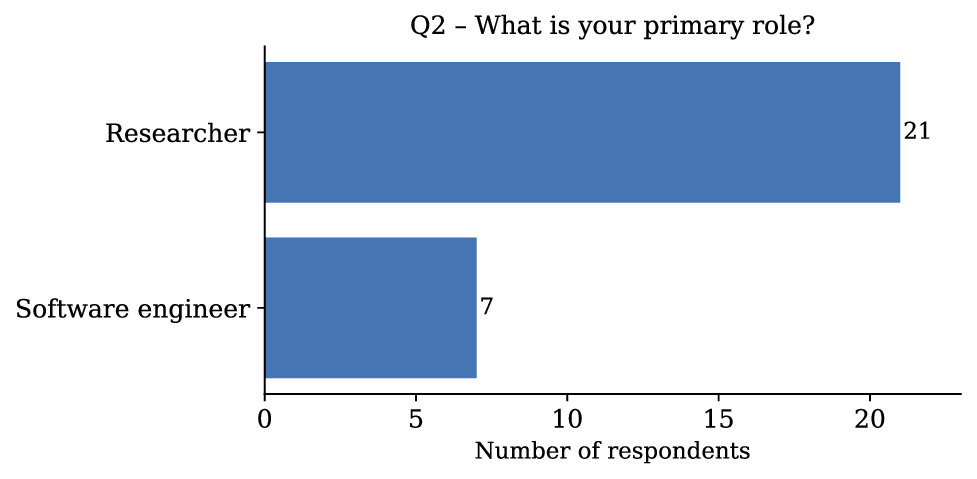}} \\
    \subfloat[Question 3.\label{fig:q3}]{\includegraphics[width=.48\columnwidth]{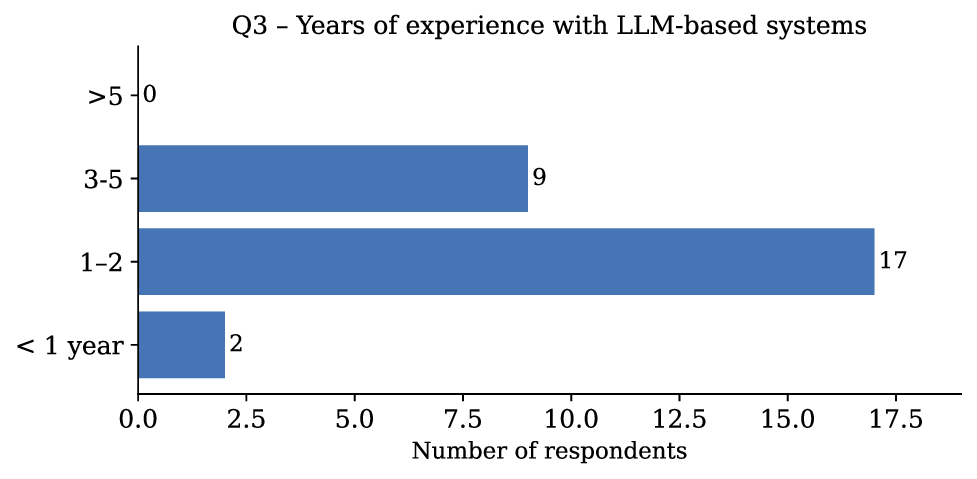}} \hfill
    \subfloat[Question 4.\label{fig:q4}]{\includegraphics[width=.48\columnwidth]{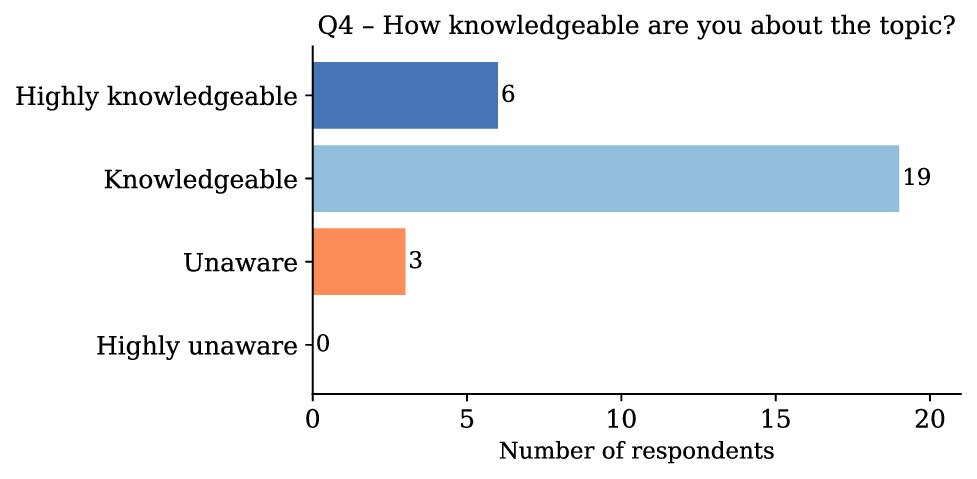}}
    \caption{Screening and background section responses.}
    \label{fig:sec1}
\end{figure}

\Cref{fig:sec1} summarises the background profile of the respondents.
All 28 participants reported having worked with \acp{llm} in software systems (Q1): $23$~($82\%$) in a research or prototyping context, and $5$~($18\%$) in production systems.
Regarding primary role (Q2), the sample comprised $21$ researchers~($75\%$) and $7$ software engineers~($25\%$).
In terms of experience with \ac{genai}ware (Q3), more than half of the respondents~($17$, $61\%$) reported $1$--$2$ years, and $9$~($32\%$) reported $3$--$5$ years; only $2$~($7\%$) had less than one year, and none reported more than five.
Regarding self-reported domain knowledge (Q4), $19$ respondents~($68\%$) described themselves as \emph{knowledgeable} and $6$~($21\%$) as \emph{highly knowledgeable}; $2$ respondents~($11\%$) reported being \emph{unaware} of the topic.
Overall, the sample reflects a technically informed population with moderate-to-high familiarity with \ac{llm}-based development.

\begin{figure}
    \centering
    \subfloat[Question 5.\label{fig:q5}]{\includegraphics[width=.48\columnwidth]{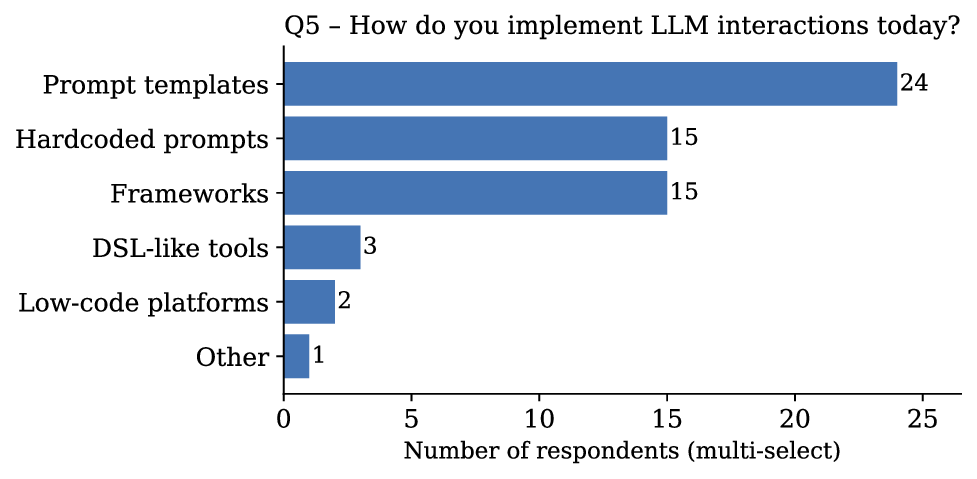}} \hfill
    \subfloat[Question 6.\label{fig:q6}]{\includegraphics[width=.48\columnwidth]{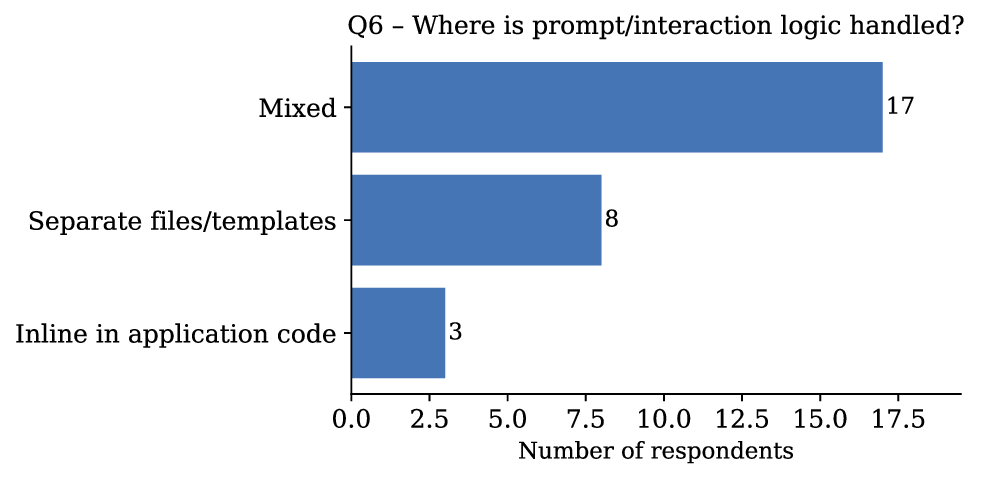}}
    \caption{Current practice section responses.}
    \label{fig:sec2}
\end{figure}

\Cref{fig:sec2} shows the responses regarding current development practices.
Prompt templates (e.g., f-strings, Jinja2) are the most widely adopted technique (Q5), used by $24$~respondents~($86\%$), followed by hardcoded prompts embedded in application code ($15$, $54\%$) and \ac{llm} orchestration frameworks such as LangChain or Semantic Kernel ($15$, $54\%$).
Only $3$~respondents~($11\%$) reported using \ac{dsl}-like tools for structured generation, two respondents~($7\%$) relied on low-code or no-code platforms, and only one respondent~($4\%$) marked other approaches.
This confirms that the use of structured, language-level abstractions remains uncommon in practice.
Regarding where interaction logic is managed (Q6), the majority ($17$, $61\%$) handle it in a mixed fashion, combining different locations; $8$~respondents~($28\%$) use separate files or templates, and $3$~($11\%$) keep it inline in application code.

\begin{figure}
    \centering
    \includegraphics[width=.72\columnwidth]{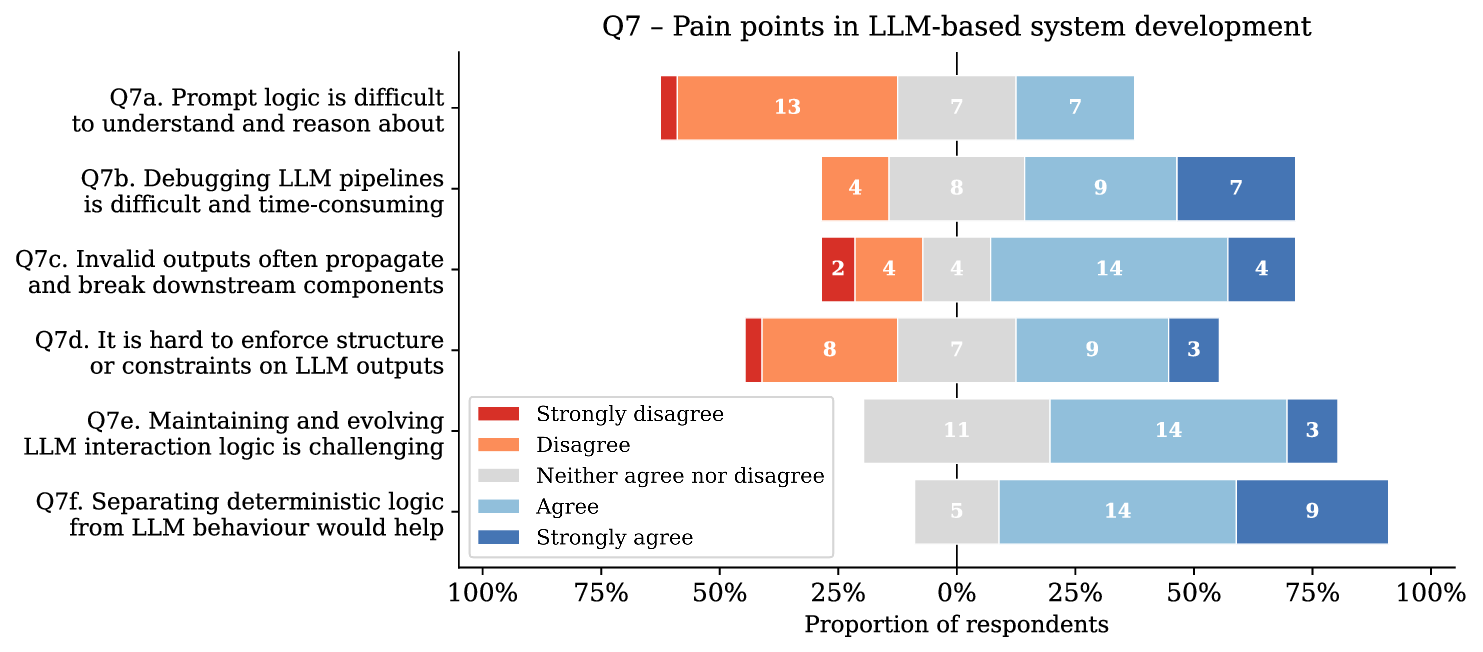}
    \caption{Pain points section responses - Question 7.}
    \label{fig:sec3}
\end{figure}

\Cref{fig:sec3} reports perceived pain points when developing \ac{genai}ware (Q7).
The strongest agreement was recorded for Q7f (\emph{Separating deterministic logic from \ac{llm} behaviour would help}; $\bar{x} = 4.14$, $\mathrm{SD} = 0.71$), with $82\%$ of respondents agreeing or strongly agreeing, directly motivating one of the core design choices of \ac{lsl}.
Q7e (\emph{Maintaining and evolving \ac{llm} interaction logic is challenging}; $\bar{x} = 3.71$) and Q7b (\emph{Debugging \ac{llm} pipelines is difficult}; $\bar{x} = 3.68$) also received predominantly positive agreement, with $61\%$ and $57\%$ of respondents rating them \emph{Agree} or \emph{Strongly agree}, respectively.
Q7c (\emph{Invalid outputs propagate and break downstream components}; $\bar{x} = 3.50$) showed similarly high agreement~($64\%$), while Q7d (\emph{Hard to enforce structure or constraints}; $\bar{x} = 3.18$) drew more mixed responses.
In contrast, Q7a (\emph{Prompt logic is difficult to understand and reason about}; $\bar{x} = 2.71$) was the only item where a majority~($50\%$) disagreed, suggesting that practitioners do not yet perceive the readability of prompt text itself as a primary challenge.

Question~8 served as an embedded attention check (see \Cref{sec:threats-mitigation}); all 20 retained respondents answered it correctly, confirming the attentiveness of the sample.

\begin{figure}
    \centering
    \includegraphics[width=.72\columnwidth]{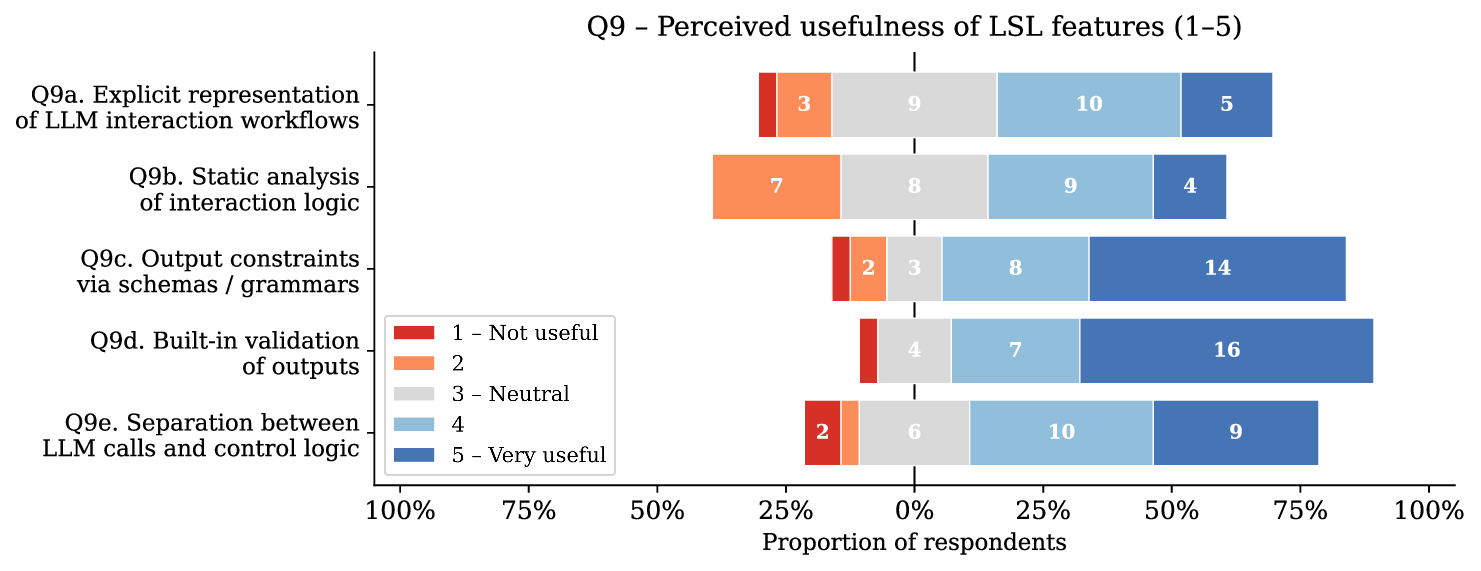}
    \caption{Desired capabilities section responses - Question 9.}
    \label{fig:sec5}
\end{figure}

\Cref{fig:sec5} shows the perceived usefulness of specific \ac{lsl} features (Q9, rated $1$--$5$ with higher values indicating greater usefulness).
Built-in validation of outputs (Q9d) received the highest mean rating ($\bar{x} = 4.32$, $\mathrm{SD} = 0.98$), with $82\%$ of respondents assigning a score of $4$ or $5$.
Output constraints via schemas or grammars (Q9c) were also rated highly ($\bar{x} = 4.14$), with $79\%$ of respondents scoring 4 or above.
The separation between \ac{llm} calls and control logic (Q9e; $\bar{x} = 3.82$) and an explicit representation of interaction workflows (Q9a; $\bar{x} = 3.54$) similarly scored above the neutral midpoint.
Static analysis of interaction logic (Q9b) received the lowest mean rating ($\bar{x} = 3.36$), though still above neutral, with $46\%$ of respondents rating it $4$ or above.
Overall, all five features were rated above neutral, indicating broad appreciation of the \ac{lsl} feature set.

\begin{figure}
    \centering
    \subfloat[Question 10\label{fig:q10}]{\includegraphics[width=.72\columnwidth]{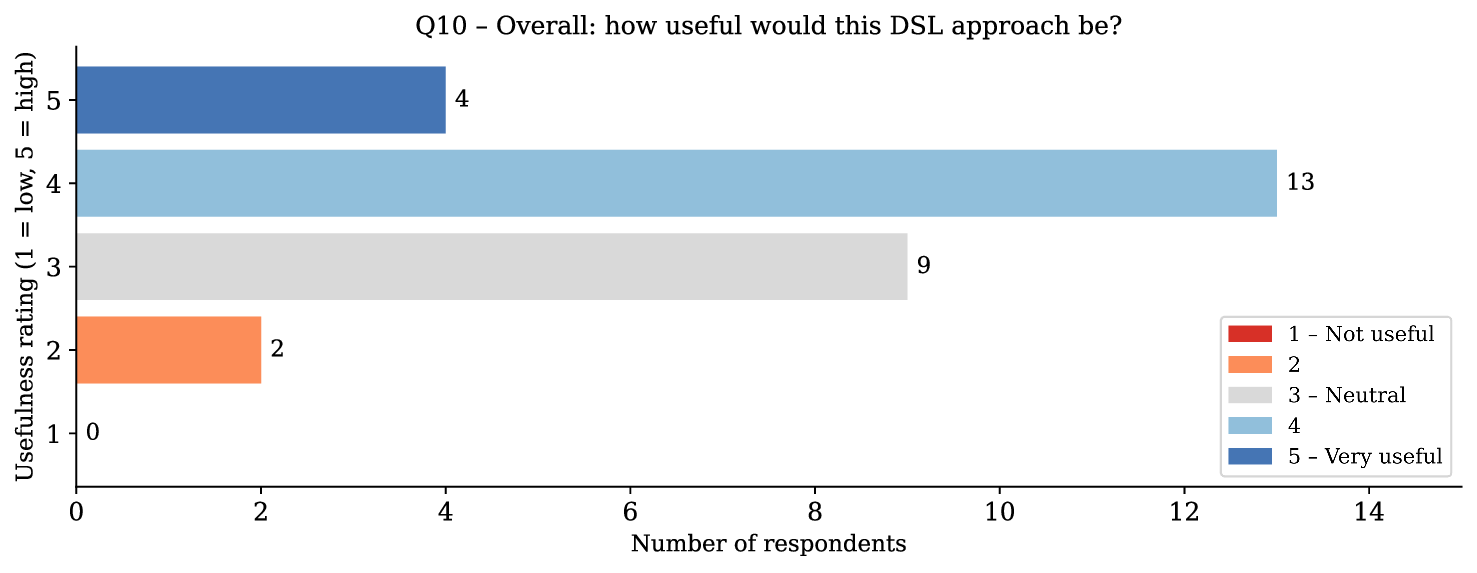}} \\
    \subfloat[Question 11\label{fig:q11}]{\includegraphics[width=.72\columnwidth]{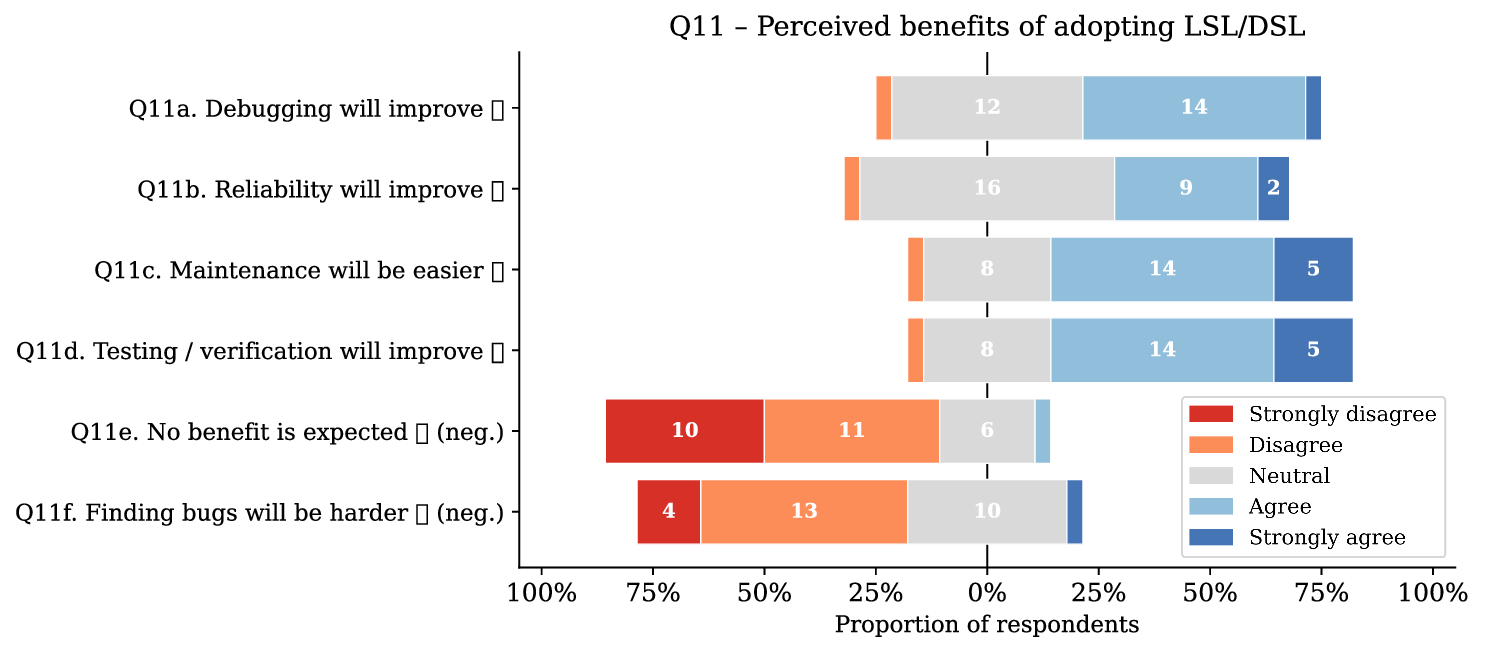}} \\
    \subfloat[Question 12\label{fig:q12}]{\includegraphics[width=.72\columnwidth]{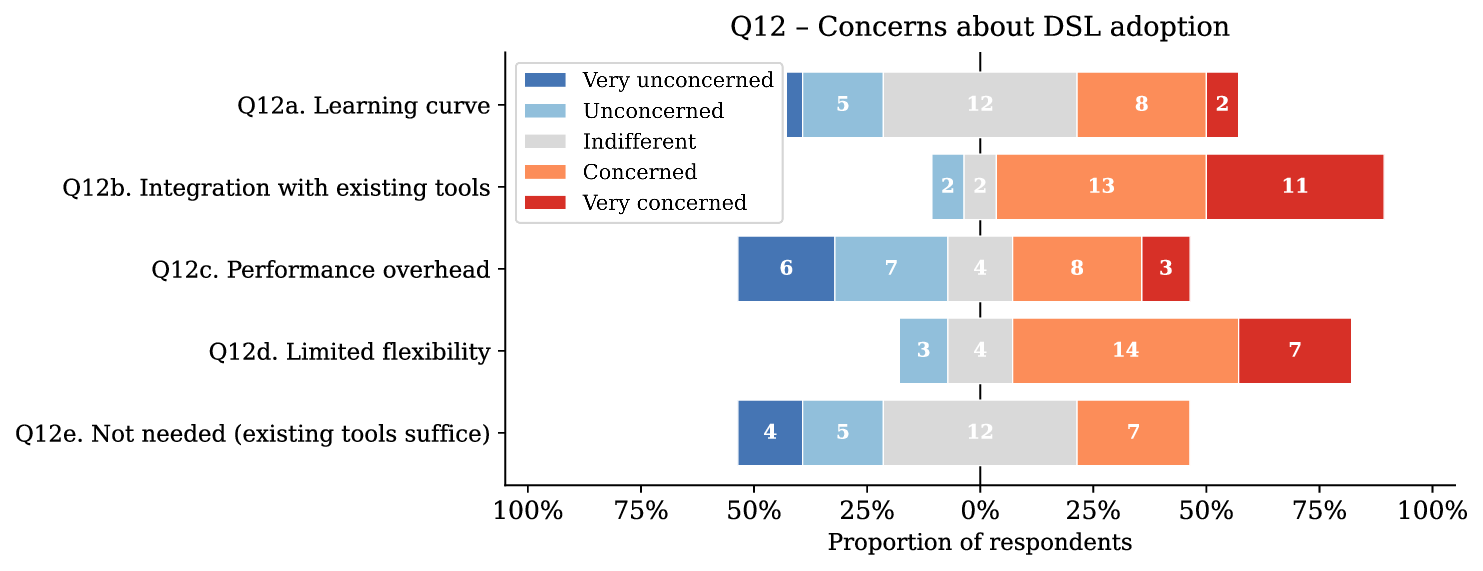}}
    \caption{Reaction to \ac{lsl} concept section responses.}
    \label{fig:sec6}
\end{figure}

\Cref{fig:sec6} reports participants' reactions to the \ac{lsl} concept.
Regarding overall perceived usefulness (Q10), the mean rating was 3.70 ($\mathrm{SD} = 0.80$, median~$= 4$), with $61\%$ of respondents assigning a score of 4 or 5 and no respondent rating the concept as not useful at all~(1).
Regarding expected benefits (Q11), respondents were broadly positive: ease of maintenance (Q11c; $\bar{x} = 3.82$) and improvements to testing and verification (Q11d; $\bar{x} = 3.82$) received the highest endorsement, with $68\%$ of respondents agreeing or strongly agreeing on both items.
Improvements to debugging (Q11a; $\bar{x} = 3.54$) were supported by $54\%$ of respondents, and reliability improvements (Q11b; $\bar{x} = 3.43$) by $39\%$.
The two negatively framed items, serving as internal consistency indicators, showed consistently low agreement: Q11e (\emph{No benefit is expected}; $\bar{x} = 1.93$) was rejected by $75\%$ of respondents, and Q11f (\emph{Finding bugs will be harder}; $\bar{x} = 2.32$) by $61\%$, confirming coherent response patterns.
Regarding adoption concerns (Q12), integration with existing tools (Q12b; $\bar{x} = 4.18$, $\mathrm{SD} = 0.86$) emerged as the primary concern, with $86\%$ of respondents reporting being \emph{concerned} or \emph{very concerned}.
Limited flexibility of the \ac{dsl} (Q12d; $\bar{x} = 3.89$) was the second most prominent concern~($75\%$ concerned or very concerned).
The learning curve (Q12a; $\bar{x} = 3.18$) and the question of whether such a language is actually necessary (Q12e; $\bar{x} = 2.79$) raised moderate concerns, while performance overhead (Q12c; $\bar{x} = 2.82$) showed the most mixed distribution of responses.

Finally, a Mann-Whitney U test comparing researchers ($n = 21$) and software engineers ($n = 7$) found no statistically significant differences across any of the four construct blocks (Q7: $U = 65.0$, $p = 0.67$; Q9: $U = 83.5$, $p = 0.61$; Q11: $U = 73.0$, $p = 1.00$; Q12: $U = 45.5$, $p = 0.14$), indicating consistent perceptions across roles; statistical power is however limited by the sample size, and these results should be interpreted accordingly.

\subsection{Threats Mitigation}
\label{sec:threats-mitigation}

To ensure the integrity of the collected data and mitigate the risk of careless responding, we adopted several established quality control mechanisms. 
Given the relatively small scale and limited length of the survey instrument, we did not apply question randomization, as fatigue and ordering effects were considered unlikely to substantially impact participant responses. 
Instead, we focused on incorporating \emph{internal consistency} and \emph{attention checks} directly into the questionnaire design and subsequent analysis.

Specifically, one of the survey items was intentionally negatively framed (with respect to another item in the same group) to allow the identification of inconsistent response patterns, such as uniformly selecting the same agreement level across all questions. 
In addition, an explicit attention check was embedded within the survey, asking participants to select a particular response option to confirm that they were carefully reading the questions. 
During data analysis, we further examined response reliability by computing the intra-individual variance across all answers; submissions exhibiting near-zero variance, indicative of repetitive or inattentive behavior, were excluded from the final dataset. 
We also assessed the internal consistency of the measured constructs using Cronbach’s alpha.

To analyze differences in perceived alignment across roles, we applied a Mann-Whitney U test.
We selected this tests because the collected responses represented only two roles, \emph{Researcher} ($n = 21$) and \emph{Software engineer} ($n = 7$), with no respondent selecting any of the remaining options.

\section{Roadmap} %
\label{sec:roadmap}

\begin{figure}[!ht]
\begin{center}
    \includegraphics[width=\columnwidth]{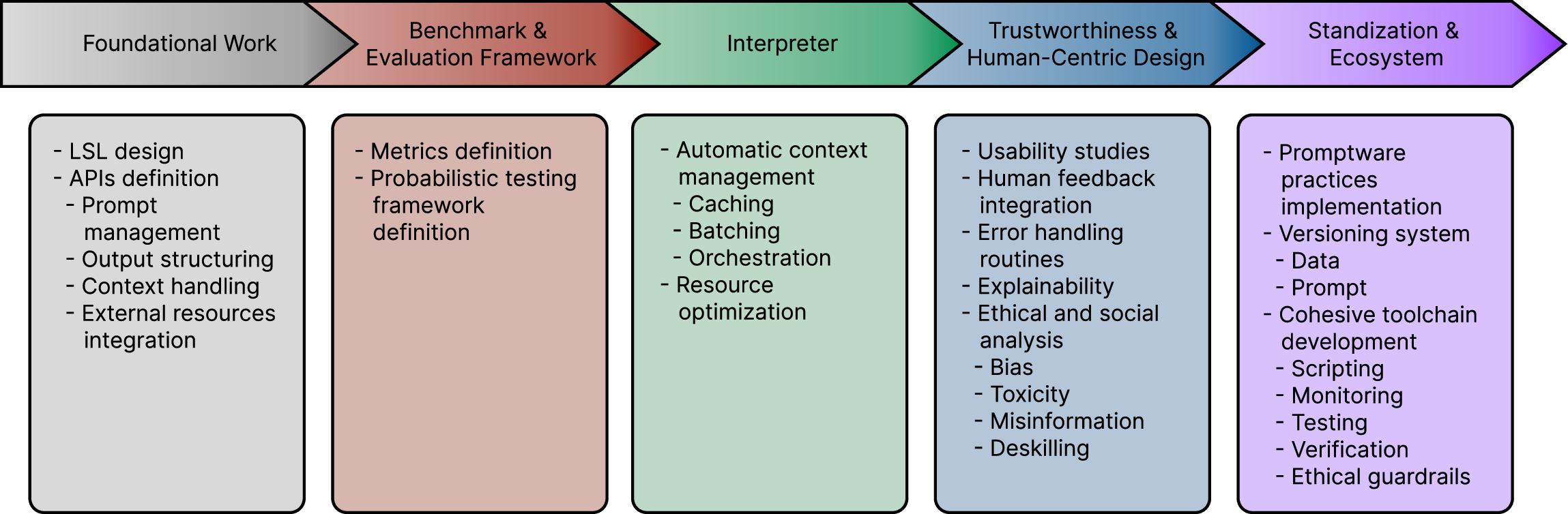}
\caption{Roadmap to the development of \ac{lsl}.}
\label{fig:roadmap}
\end{center}
\end{figure}

Building on the opportunities identified in \Cref{sec:opportunities} and the challenges outlined in \Cref{sec:challenges}, we envision a broad research roadmap towards reliable, robust, and trustworthy \ac{genai}ware. %
This roadmap, visualised in \Cref{fig:roadmap}, outlines research directions for the community at large and highlights how language design, API development, interpreter infrastructures, usability, and ethical oversight must come together.

\paragraph{Foundational Work}
The development of languages and abstractions will allow for structured interactions with \acp{llm}.
This includes the design of domain-specific languages such as \ac{lsl}, which can provide formalisms for scripting, constraining, and reasoning about LLM interactions.
The community will need to explore various levels of abstraction, ranging from low-level APIs for fine-grained control to high-level declarative interfaces that prioritize usability and accessibility.
Alongside the language itself, the definition of open and standardized APIs is critical.
Such interfaces should unify access to prompt management, structured outputs, context handling, and integration with external resources such as datasets or knowledge bases, thereby fostering interoperability and extensibility.

\paragraph{Benchmarks and Evaluation Frameworks}
Improving the reliability and robustness of \ac{genai}ware %
requires a shift from ad-hoc testing to systematic evaluation practices.
Research is needed to define benchmarks and metrics that extend beyond model accuracy and capture safety, correctness, and alignment with user expectations.
Evaluation should not be limited to \acp{llm} in isolation but applied to the pipelines and applications in which they are embedded.
Probabilistic testing, adversarial evaluation, and methods for runtime validation of generated outputs are especially promising directions.
By establishing common benchmarks and frameworks, the community can enable more rigorous comparisons across tools and systems, foster reproducibility, and accelerate progress.

\paragraph{Interpreters}
Interpreter infrastructures form the backbone of scripted \ac{llm} interactions and require dedicated research attention.
Efficient interpreters should be capable of managing context automatically, exploiting opportunities for caching and batching, and orchestrating parallel execution where tasks allow.
They must also balance usability and performance, providing developers with intuitive abstractions while optimizing resource utilization under the hood.
This line of research naturally connects with the design of compiler and runtime systems, yet introduces unique challenges related to the causal nature of \acp{llm} and the integration of external resources.
An interpreter that is both efficient and transparent will be decisive in making scripted \ac{llm} systems practical at scale.

\paragraph{Trustworthiness \& Human-centric Design}
Usability studies, mechanisms for domain experts to shape system behavior, and frameworks for ensuring meaningful human control should become an integral part of \ac{genai}ware. %
Human feedback must not be treated as an afterthought, but rather be integrated systematically to guide error handling and evaluation.
Explainability tools should be embedded into workflows to allow developers and users to trace outputs, debug prompts, and build confidence in the system's behavior.
At the same time, ethical and social concerns like bias, toxicity, misinformation, deskilling, and job displacement require continuous monitoring and proactive mitigation strategies.
Addressing these aspects is essential to ensure trustworthiness and long-term societal acceptance.

\paragraph{Standardisation \& Ecosystems}
Standardisation represents another key frontier.
Just as MLOps practices evolved to bring order to machine learning pipelines, the community now needs standards for promptware, prompt and data versioning, and lifecycle management of \ac{genai}ware. %
Without such common ground, development practices will remain fragmented and inconsistent.
Beyond formal standards, an interoperable ecosystem of tools must emerge.
Instead of today's patchwork of isolated frameworks, we need cohesive toolchains that integrate scripting, monitoring, testing, verification, and ethical safeguards into the everyday workflows of developers.

\smallskip
Beyond this roadmap, our long-term vision is to embed \ac{llm} scripting and interaction into the broader DevOps lifecycle.
In such a future, prompts, data, and scripts will be treated as first-class citizens alongside code, benefiting from the same rigor in versioning, testing, and verification.
Verification, monitoring, and human oversight will be embedded throughout development and deployment pipelines, ensuring that reliability and transparency are not afterthoughts but built-in guarantees.
This vision requires the construction of sustainable ecosystems in which \ac{genai}ware %
can be designed, deployed, and maintained with the same degree of trust, robustness, and societal accountability as traditional software.

\section{Conclusion}
\label{sec:conclusion}

\acp{llm} represent valuable assets to build \ac{ai}-powered applications.
Nevertheless, they come with several risks and limitations that prevent to freely adopt them into fully automated pipelines and workflows.
Throughout this paper, we have argued that \ac{se} plays a decisive role in addressing these limitations.

The core idea is to develop a \ac{dsl} that works as a scripting language, namely \ac{lsl}, to automate the templating of the input prompt and the constraining of the generated completions of the \ac{llm} to enforce guarantees on the \ac{llm} behaviour.
An interpreter complements \ac{lsl} in multiple ways, like
\begin{enumerate*}
    \item managing the \ac{llm} context,
    \item managing the interfaces towards the model and the external resources,
    \item managing error correction subroutines,
    \item running consistency checks, and
    \item producing explanations when needed.
\end{enumerate*}

Building on that, we outlined a roadmap that spans foundational directions in language and API design, the development of benchmarks and evaluation frameworks, the creation of efficient interpreter infrastructures, the integration of human-centric feedback and usability studies, and the establishment of standards and ecosystems that can bring order to today’s fragmented practices. This agenda emphasizes that progress requires not only technical innovations but also attention to social, ethical, and organizational challenges, so that trustworthiness, transparency, and societal accountability are built into \ac{llm}-powered systems (\ac{genai}ware) from the outset.

Our survey study, conducted with $28$ practitioners and researchers experienced in \ac{genai}ware development, provides initial empirical support for the relevance of \ac{lsl}.
Respondents broadly confirmed the practical pain points targeted by \ac{lsl}: $82\%$ agreed that separating deterministic logic from \ac{llm} behaviour would be beneficial~(Q7f; $\bar{x} = 4.14$), and maintaining and debugging \ac{llm} interaction logic were rated as challenging by a clear majority.
All five proposed \ac{lsl} features were perceived as useful above the neutral midpoint, with built-in output validation and constraint enforcement receiving the highest ratings~($\bar{x} = 4.32$ and $4.14$, respectively).
Overall, $61\%$ of respondents rated the \ac{dsl} concept as useful or very useful (score~$\geq 4$ on a 1--5 scale; $\bar{x} = 3.68$), while integration with existing toolchains and language flexibility emerged as the primary adoption concerns to be addressed in future work.

\section*{Data Availability}

The replication package of our experiments is available at \url{https://github.com/vincenzo-scotti/tosem_roadmap_to_2030}.

\begin{acks}
Some parts of our work were assisted by Generative AI. 
We used Anthropic's Claude Opus 4.7 and Microsoft's Copilot (based on OpenAI's ChatGPT 5 model family) to help us refine the manuscript and prepare the visualization of the evaluation outcome.

This work was supported by funding from the pilot program Core Informatics at KIT (KiKIT) of the Helmholtz Association (HGF) and KASTEL Security Research Labs, Karlsruhe.
\end{acks}

\bibliographystyle{ACM-Reference-Format-num}
\bibliography{bibliography}

\end{document}